\documentclass[12pt]{article}

\input psfig

\def\xslide#1#2#3#4#5#6#7{\centerline{\psfig
{figure=#1,height=#2,bbllx=#3bp,bblly=#4bp,bburx=#5bp,bbury=#6bp,width=#7,clip=}}}

\title{Scalar Mesons and Multichannel Amplitudes}
\author{R. Kami\'nski$^1$, L. Le\'sniak$^1$ and B.\ Loiseau$^2$\\ \\
{\em $^1$Department of Theoretical Physics}\\
{\em H. Niewodnicza\'nski Institute of Nuclear Physics,}\\ 
{\em PL 31-342 Krak\'ow, Poland}\\
{\em $^2$LPTPE Universit\'e P. et M. Curie, 4, Place Jussieu,} \\
{\em              75252 Paris CEDEX 05, France}}

\newcommand {\bfk}{{\bf k}}

\newcommand {\be}{\begin{equation}}
\newcommand {\eb}{\end{equation}}
\newcommand{\ba}{\begin{eqnarray}}
\newcommand{\ea}{\end{eqnarray}}
\newcommand{\pp}{$\pi\pi$ }
\newcommand{\roro}{$\sigma\sigma$ }
\newcommand{\kk}{$K\overline{K}$ }
\newcommand{\fo}{$f_0(980)$ }
\newcommand{\epw}{$f_0(1400)$ }
\newcommand{\epsig}{$f_0(500)$ }
\newcommand{\fgg}{$f_0(1500)$ }
\newcommand{\fj}{$f_0(1710)$ }
\newcommand{\reactpol}{$\pi^- p_{\uparrow} \rightarrow \pi^+ \pi^- n$ }
\newcommand{\reactnon}{$\pi^- p \rightarrow \pi^+ \pi^- n$ }

\newcommand{\sig}{$\sigma$ }
\newcommand{\vs}{\vspace{0.2cm}}

\begin{document}

 \maketitle

hep-ph/9810386

\begin{abstract}

Properties of scalar--isoscalar mesons are analysed in an unitary
model using separable interactions in three decay channels: \pp,
\kk and an effective $2\pi2\pi$. 
We obtain different solutions by fitting various data on the 
\pp and \kk phase shifts and
inelasticities including the CERN-Cracow-Munich measurements of the
\reactpol reaction on a polarized target. Ana\-ly\-ti\-cal structure of the
meson-meson multichannel amplitudes is studied with a special emphasis
on the role played by the $S$-matrix zeroes.  $S$-matrix poles,
located in the complex energy plane not
too far from the physical region, are interpreted as
scalar resonances. We see a wide
\epsig, a narrow \fo and a relatively narrow \epw. 
In one of our solutions a resonance
at about 1700 MeV is also found. Total, elastic and inelastic channel
cross sections, branching ratios and coupling constants are evaluated and
compared with available data. We construct  an  approximation to our model  and show that
the Breit-Wigner approach has a limited phenomenological applicability.
\end{abstract}

%-------------------------------------------------------------
\section{Introduction}

\hspace{0.7cm}
Full classification and identification of scalar mesons is not yet well
established 
\cite{pdg98}.
Many theoretical and experimental efforts have been recently made for a
better understanding  of the scalar mesons
as it can be seen, for example, in references given in 
\cite{pdg98,hadron97}. 
From QCD one expects presence of some scalar
($J^{PC} = 0^{++}, I=0$) glueballs which can be mixed with ordinary
$q\overline{q}$ scalar states \cite{close97}.  The lowest scalar
glueball masses predicted  by lattice QCD calculations are in the range
1500 MeV to 1700 MeV \cite{weingarten, michael}.  There are now lively
discussions about the nature of scalar mesons $f_0(1500)$ and
$f_J(1710)$ and their possible mixture with scalar glueballs
\cite{weingarten, close98}.  High statistics experiments on meson
productions such as $p\overline{p}$ annihilation, $\pi N$ scattering
on unpolarized and polarized targets, central production in
$pp$ scattering, $J/\Psi$
or other heavy meson decays and $\gamma\gamma$ collisions have been
performed.  Their analyses give some new evidence for the existence of
five scalar-isoscalar mesons $f_0(400-1200)$ or $\sigma$, \fo,
$f_0(1370)$, $f_0(1500)$ and $f_0(1710)$ \cite{pdg98}.

In \cite{kll} we have analysed new solutions for
the scalar-isoscalar \pp phase shifts \cite{klr} together with
previous \kk results in the framework of a three coupled channel model
based on an extension of the two-channel model of \cite{klm}.
Separable potentials were used to describe interactions in the \pp,
\kk and an effective $2\pi2\pi$ channels. 
Let us recall that use of separable potentials leads to an analytical
solution for the $S$-matrix which helps to check and understand the results.
It furthermore represents one of the easiest way to handle non-local
interactions and so it should be adequate to describe, with a minimum number
of parameters, strong energy dependence in several coupled channel reactions
as it seems to be the case here. Simple analytical form of separable
interactions is then phenomenologically very useful in fitting the \pp
and \kk experimental data from the corresponding thresholds up to 1.8 GeV.

The third effective $2\pi2\pi$ channel was
introduced to take into account a strong four-pion production observed
in different experiments. In these data some evidence for four pion
clustering into \roro pair, coming from strong interaction between two
pions was found (see for example \cite{abele96}). Our $2\pi2\pi$
effective channel, called \roro, can however represent also other
possible clusterings such as $\rho\rho$, 
$a_1(1260)\pi$, $\pi(1300)\pi$
and $\omega\omega$. One should not mix the effective threshold mass
with twice the mass of the \epsig or $\sigma$ resonance which is seen in 
the \pp channel.

The parameters of the model were determined by a fit to two sets of \pp
phase-shifts and inelasticities obtained in a recent analysis of the CERN-
Cracow-Munich measurements of the $\pi^-p_\uparrow
\longrightarrow\pi^+\pi^-n$ reaction on a {\it polarized} target
\cite{klr}. 
It was stressed in 
\cite{klr} that the $a_1$ exchange gives an important contribution to the
\reactnon
reaction amplitudes.
Recently Achasov and Shestakov have also come to the conclusion that the $a_1$ exchange 
plays an important role in the reaction 
$\pi^- p \rightarrow \pi^0\pi^0n$ on unpolarized target
\cite{achasov}.
This conclusion may be experimentally verified by {\em measurements on polarized target}.   
The \pp data
\cite{klr}, 
covering the 600 to 1600 MeV \pp invariant mass
range, were completed in the lower energy region by the \pp phase shifts of 
\cite{rosselet,belkov} 
and 
\cite{srinivasan}. 
Further constraints were
imposed by using the \kk phase shifts from the \kk threshold up to 1530 MeV 
\cite{cohen}.

We found a relatively narrow (90-180 MeV) scalar resonance
$f_0(1400-1460)$. Our analysis of previous CERN-Munich unpolarized target
data 
\cite{grayer} 
predicted a much broader ($\Gamma\approx$ 500 MeV) state.
We have also obtained a very wide ($\Gamma\approx$ 500 MeV) $f_0(500)$
resonance and the well established narrow $f_0(980)$ ($\Gamma\approx$ 60-70
MeV). 
Our model allows a theoretical study on the origin of resonances by
switching off the interchannel couplings. In all the solutions found in
\cite{kll} 
the \kk interaction was repulsive or not attractive enough to
create, by itself, a \kk bound state. 

In this paper we built two new
solutions in which the \kk state is bound in the uncoupled case. Our
solutions are also characterized by presence or absence of a \roro
bound state. Gradual increase of interchannel couplings allows one to link
the S-matrix poles from the uncoupled to fully coupled case. In our
approach experimental resonances correspond to the S-matrix poles in the
complex energy plane close to the physical region. This is in contrast to
some other descriptions where resonance parameters are introduced using
K-matrix poles or Breit-Wigner formulae with some ad-hoc background in
each channel. Let us here remark that the Particle Data Group 
\cite{pdg98}
has misplaced the \sig meson parameters found in 
\cite{klm}. 
They have been
referred to under the name"$f_0(400-1200)$ Breit-Wigner mass or K-matrix
pole parameters" on page 363 of 
\cite{pdg98} 
instead of being referred under
the appropriate title "$f_0(400-1200)$ T-matrix pole $\sqrt s$".
Similar
misplacements were also made on pages 392 and 393 for the $f_0(1370)$.
Studying the analytical structure of multichannel amplitudes we shall
show the important role played not only by poles but also by zeroes of the
S-matrix. Knowledge of the poles and zeroes enables us to give in some
cases a simple phenomenological approximation of the T-matrix. Our results
on scalar-isoscalar resonances, masses, widths, branching ratios,
coupling constants as well as phase shifts, inelasticities and cross
sections in the three channels will be discussed for our different
solutions and compared with the available data. 

In Section 2 we describe the two new solutions in addition to those previously presented in
\cite{kll}.  
Section 3 contains an analysis of positions of the $S$-matrix poles in the three channel
amplitudes.
Section 4 describes the influence of the $S$-matrix
poles and zeroes on phase shifts and inelasticities near the \epw resonance.
In Section 5 we comment on the limited applicability of Breit-Wigner approach in
the multichannel meson scattering.
Evaluation of branching ratios in two and three coupled channels is presented 
in Section 6 and coupling constants are discussed in Section 7.
In Section 8 we present phenomenological parametrizations of multichannel amplitudes.
Summary and conclusions are given in Section 9.
In Appendix A the full formula for the Jost function of our three-channel 
model is supplied.
Approximated formulae for the pion-pion $S$-matrix element, 
especially useful in vicinity of the \epw resonance,  
are given in Appendix B. 
  
%%%%%%%%%%%%%%%%%%%%%%%%%%%%%%%%%%%%%%%%%%%%%%%%%%%%%%%%%%%%%%%%%%%%%%%%%%%%%%%%%%%%%%
  
\section{New and former solutions}
 
\hspace{0.7cm} 
In 
\cite{kll}
we have briefly presented our three-channel model of meson-meson scattering.
This model describes simultaneously nine reactions in a unitary way.
In addition to the \pp and \kk channels an effective \roro channel is introduced
in order to describe important $4\pi$ production and rescattering processes.
Here we use exactly the same notation and definitions as in 
\cite{kll}. We furthermore 
 give, in Appendix A, the full formula for the Jost function $D(k_1,k_2,k_3)$.
Its knowledge is sufficient to construct all the $S$-matrix elements and all the 
physical quantities which will be discussed in this paper.
In
\cite{kll}
we have obtained four three-channel solutions called A, B, C and D
based on $\chi^2$ fits to the \pp $S$-wave isoscalar phase shifts and 
inelasticities, and to the \kk phase shifts.
The parameters of the separable potentials, for all the different solutions, 
have been obtained by fitting the results of the fully coupled channel
calculations to the experimental data.
If the interchannel couplings are switched off then the 
\kk pair remains unbound in all four fits.
In solutions B, C and D the \kk potential is repulsive,
so a \kk bound state cannot be formed.
In the fit A the \kk potential is attractive but its strength is too small to form
a bound state.
Parameters of the \kk interactions, found in
\cite{kll},
were different from the \kk parameters obtained in 
\cite{klm}
where a \kk bound system could be formed in absence of the interchannel interactions.
We should, however, remember that in 
\cite{klm} 
a different \pp phase-shift solution, obtained from the data analysis on an unpolarized target by the 
CERN-Munich group
\cite{grayer},
was used.

In 
\cite{kll} we have used several sets of the \pp phase shifts obtained in
\cite{klr}
from the data taken on a polarized target by the CERN-Cracow-Munich collaboration
\cite{becker}.
Now, one can ask if the data of 
\cite{klr}
rule out any set of potential parameters leading to a \kk bound state.
 We have therefore repeated the $\chi^2$ fits performed in 
\cite{kll} for the "down-flat" \pp solution by adding an additional constraint, namely that
the \kk potential
is sufficiently strong to obtain a bound state in the uncoupled case.
As a result of these studies we have obtained two new solutions E and F which will be 
described below.
Similar studies could be done for the "up-flat" solutions C and D which have less
good $\chi^2$ values as seen in Table 2 of
\cite{kll}.
%#####################################################################################
\begin{table}[h]
\centering
\caption{Separable potential parameters for the solutions E
and F. Values of $\beta$ and $m_3$ are given in GeV.}

\vspace{.7cm}

\begin{tabular}{|l|c|c|}
\hline
parameter & Solution E & Solution F \\
\hline
$\Lambda_{11,1}$ &   $-.26349\times10^{-3}$ & $-.13678\times10^{-3}$ \\
$\Lambda_{11,2}$ &   $-.18316$              & $-.17845$              \\
$\Lambda_{22}$   &   $-.60400$              & $-.52087$              \\
$\Lambda_{33}$   &   $.17703\times 10^{2}$  & $-.73962$              \\
$\Lambda_{12,1}$ &   $ .28776\times 10^{-4}$& $.89713\times10^{-6}$  \\
$\Lambda_{12,2}$ &   $.036838$ & $.048444$  \\
$\Lambda_{13,1}$ &   $-.34811\times 10^{-3}$& $-.14046\times10^{-3}$ \\
$\Lambda_{13,2}$ &   $.55929$               & $.095244$  \\ 
$\Lambda_{23}$   &   $-1.5951$     & $-.024116$ \\
$\beta_{1,1}$    &   $.16355\times 10^4$    & $.31518\times 10^4$    \\
$\beta_{1,2}$    &   $1.1052$      & $1.0712$      \\
$\beta_2$        &   $1.4960$      & $2.1224$      \\
$\beta_3$        &   $.092701$ & $1.4958$      \\
$m_3$            &   .675                   &.680                    \\
\hline
\end{tabular}
\label{parame}
\end{table}
\begin{table}[h]
\centering
\caption{Comparison of four solutions fitted to the "down-flat" data
of   [8]. Second and third rows indicate  presence or absence
of bound states in \kk and \roro channels when interchannel couplings
are switched off. In the remaining rows the different $\chi^2$ values are
specified.
Numbers of experimental points are indicated in parenthesis. 
The $\overline{\chi^2}$ values result from fitting with reduced $\eta$ errors
as explained in [7].}

\vspace{0.5cm}

\begin{tabular}{|c|c|c|c|c|}
\hline
Solution & A & B & E & F \\
\hline
bound \kk & no & no & yes & yes \\
\hline
bound \roro & no & yes & no & yes \\
\hline
$\chi^2_{\pi}$ (65) & 63.0 & 61.2 & 64.8 & 65.4 \\
\hline
$\chi^2_{\pi K}$ (21) & 15.9 & 9.7 & 17.7 & 24.7 \\
\hline
$\chi^2_{\eta}$ (30) & 13.2 & 12.9 & 12.6 & 10.5 \\
\hline
$\chi^2_{tot}$ (116) & 92.1 & 83.8 & 95.1 & 100.6 \\
\hline
& & & & \\
$\overline{\chi^2_{\eta}}$ (30) & 36.7 & 29.3 & 33.4 & 33.6 \\
\hline
& & & & \\
$\overline{\chi^2_{tot}}$ (116) & 115.6 & 100.1 & 115.9 & 123.7 \\
\hline
\end{tabular}
\label{chi}
\end{table}

The resulting potential parameters for the solutions E and F are presented in Table \ref{parame}.
For the solution E we can notice a particularly strong repulsive coupling constant $\Lambda_{33}$ in the \roro
channel. 
So the \roro interaction for the solution E will not form any bound state when 
the interchannel couplings are  set equal to zero.
This is in contrast to the solution B where such a state exists. 
The \roro bound state is also present in the solution F which has in addition a bound \kk
state.
In Table \ref{chi} we give the binding properties when couplings between channels are switched off and the 
different $\chi^2$ values of the four solutions A, B, E and F.
The $\chi^2$ values for the solution F are not as good as for the solutions A and B but 
they are still quite reasonable.
The \pp phase shifts of solution F increase by almost $180^o$ at $E \approx 1350$ MeV
since at this energy a very narrow resonance ($\Gamma \approx 0.5$ MeV) is created 
about 10 MeV below the \roro threshold.
This resonance makes also very narrow structures in the energy dependence of
the \pp inelasticity and phase shifts.
They cannot be, however, uniquely confirmed by the existing data, so we tend to treat
the solution F as an interesting but not well experimentally confirmed example of
a phenomenological set of separable potential parameters.
The $\chi^2$ numbers of the solution E are better than those of the 
solution F.
They are comparable to those of the solution A, however, the solution B
has the best $\chi^2$.
Therefore in the following mainly the three solutions A, B and E will be simultaneously 
discussed.
We shall furthermore present new physical quantities like cross sections, branching ratios and coupling
constants which have not been discussed in paper
\cite{kll}.

In Fig. 
%\ref{etae} 
1
we compare, for  energies above 1350 MeV (\roro threshold), the inelasticities for 
the solution E in the three channels.
Both \pp and \kk inelasticities show a minimum with a different depth
near 1600 MeV and 1650 MeV, respectively.
The \roro inelasticity, however, has two minima at 1475 and 1675 MeV.
Second minimum can be related to an additional scalar resonance at about
1700 MeV (\fj). 
Appearance of this resonance in addition to \epw is a unique feature
of the solution E.
No such state exists in solutions A and B as it will be shown in detail in the next 
chapter.
As we can see in Fig. 
%\ref{ephase} 
2
this state does not produce any strong increase
of the \pp phase shifts which is often expected in the vicinity of a resonance.
This fact will be explained in the next section.

%%%%%%%%%%%%%%%%%%%%%%%%%%%%%%%%%%%%%%%%%%%%%%%%%%%%%%%%%%%%%%%%%%%%%%%%%%%%%%

\begin{figure}[ptb]
    \begin{center}
\xslide{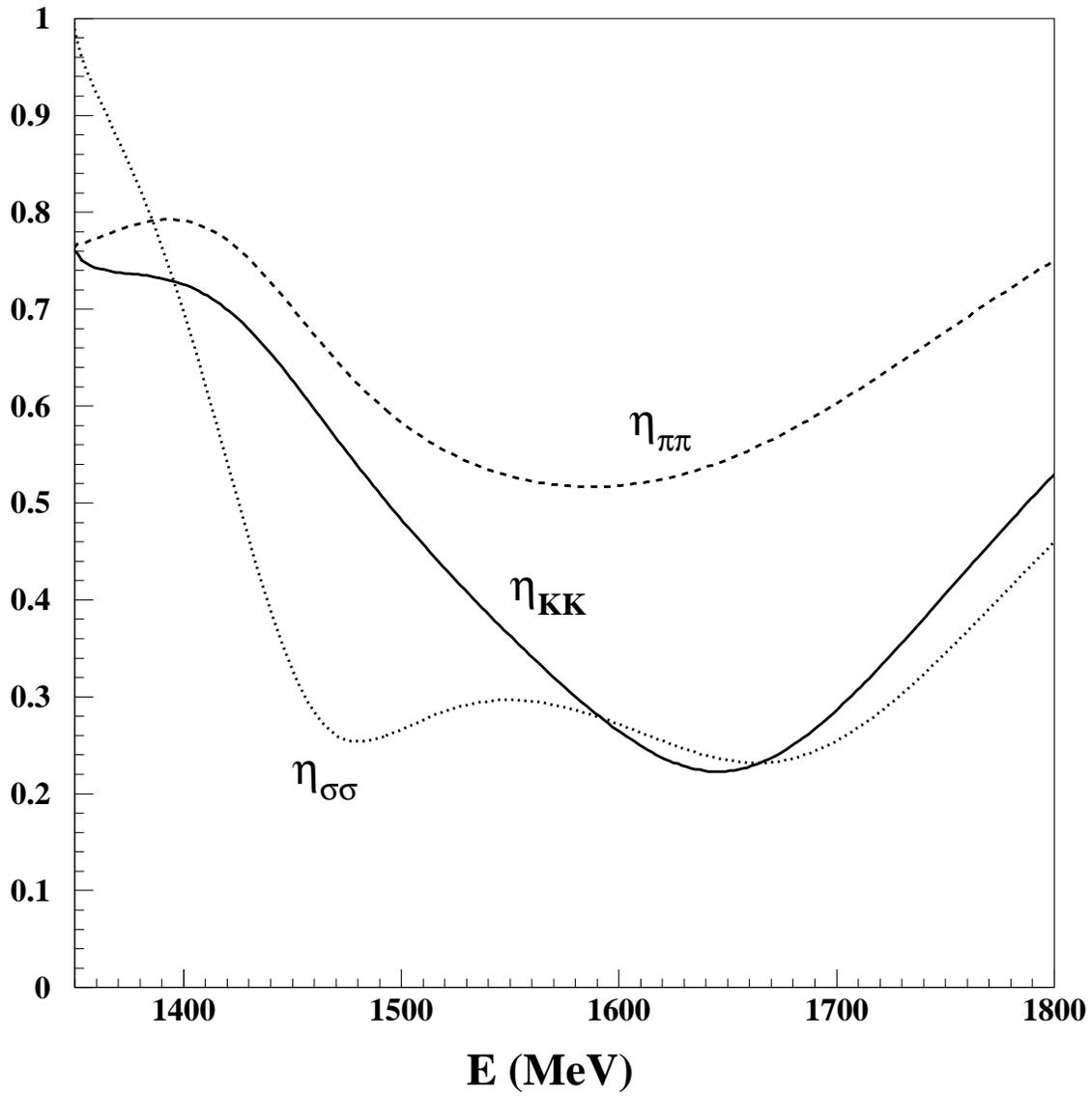}{15cm}{40}{150}{535}{650}{15cm} 
\caption{Energy dependence of inelasticity in the \pp, \kk and \roro
channels for the solution E} 
    \end{center}
                \label{etae} 
  \end{figure}

\begin{figure}[ptb]
    \begin{center}
\xslide{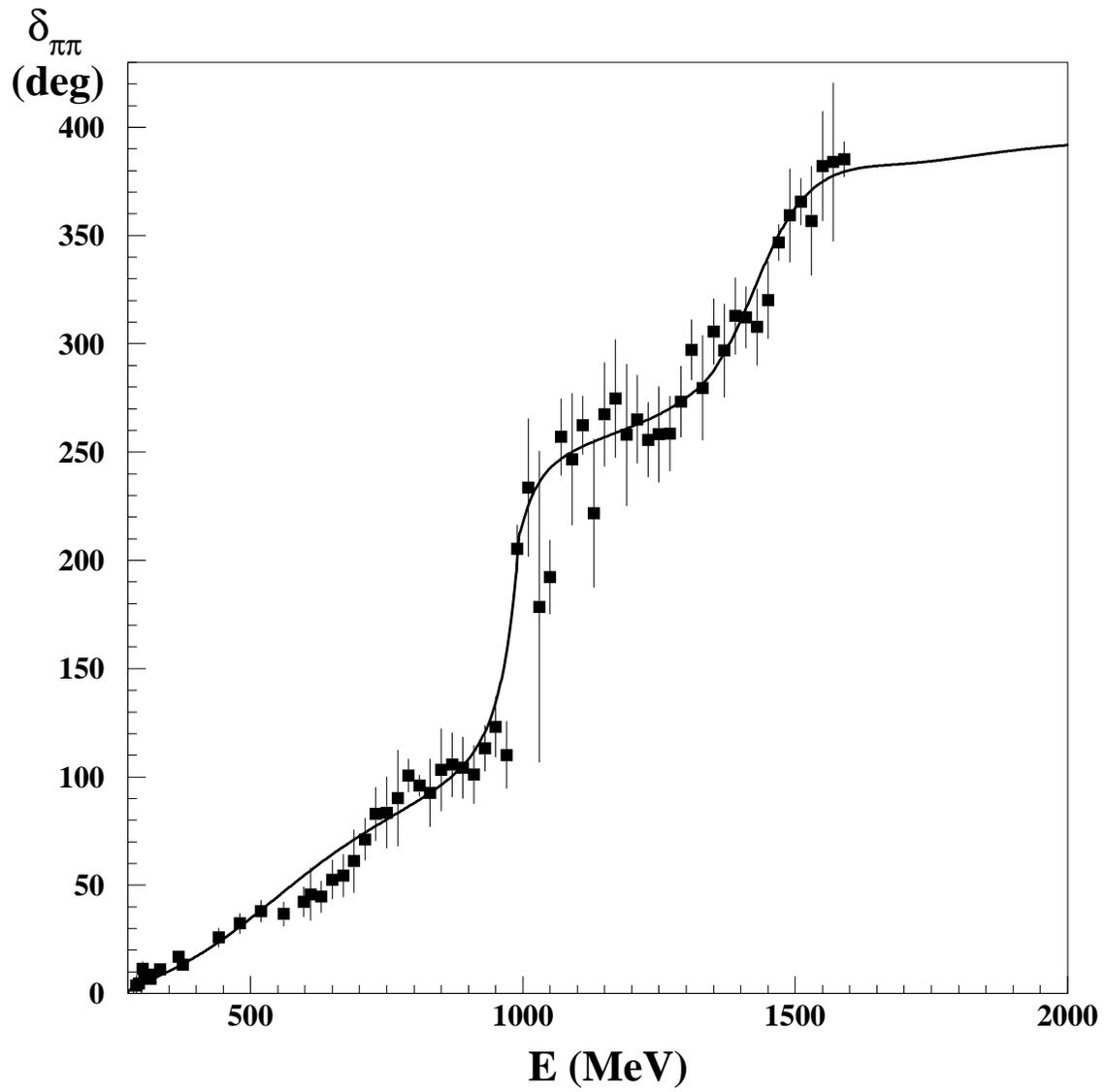}{15cm}{17}{150}{535}{672}{15cm} 
\caption{Energy dependence of \pp phase shifts for the solution E} 
    \end{center}
                \label{ephase} 
  \end{figure}

\begin{figure}[ptb]
    \begin{center}
\xslide{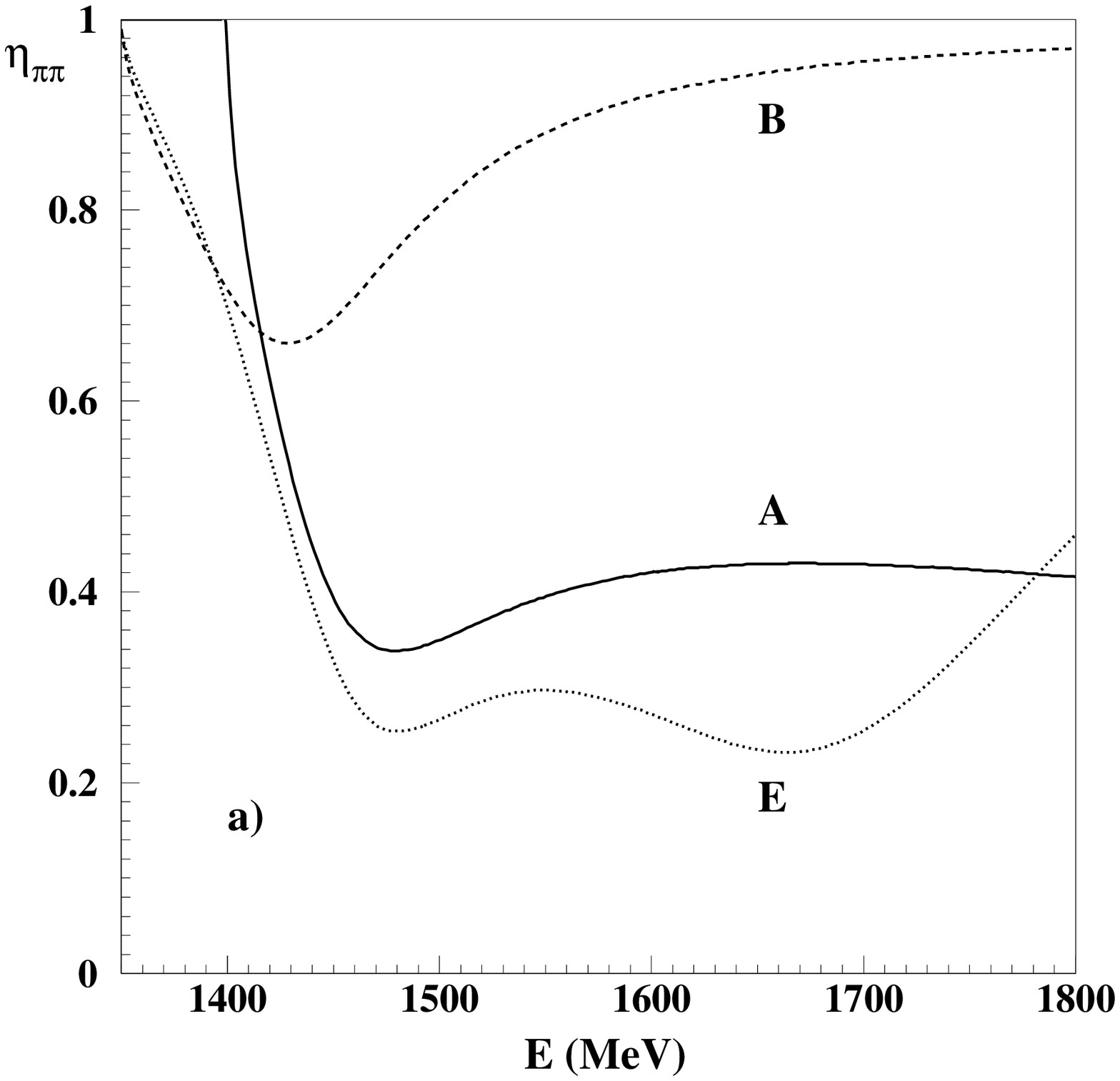}{8cm}{17}{150}{535}{665}{9cm} 
\xslide{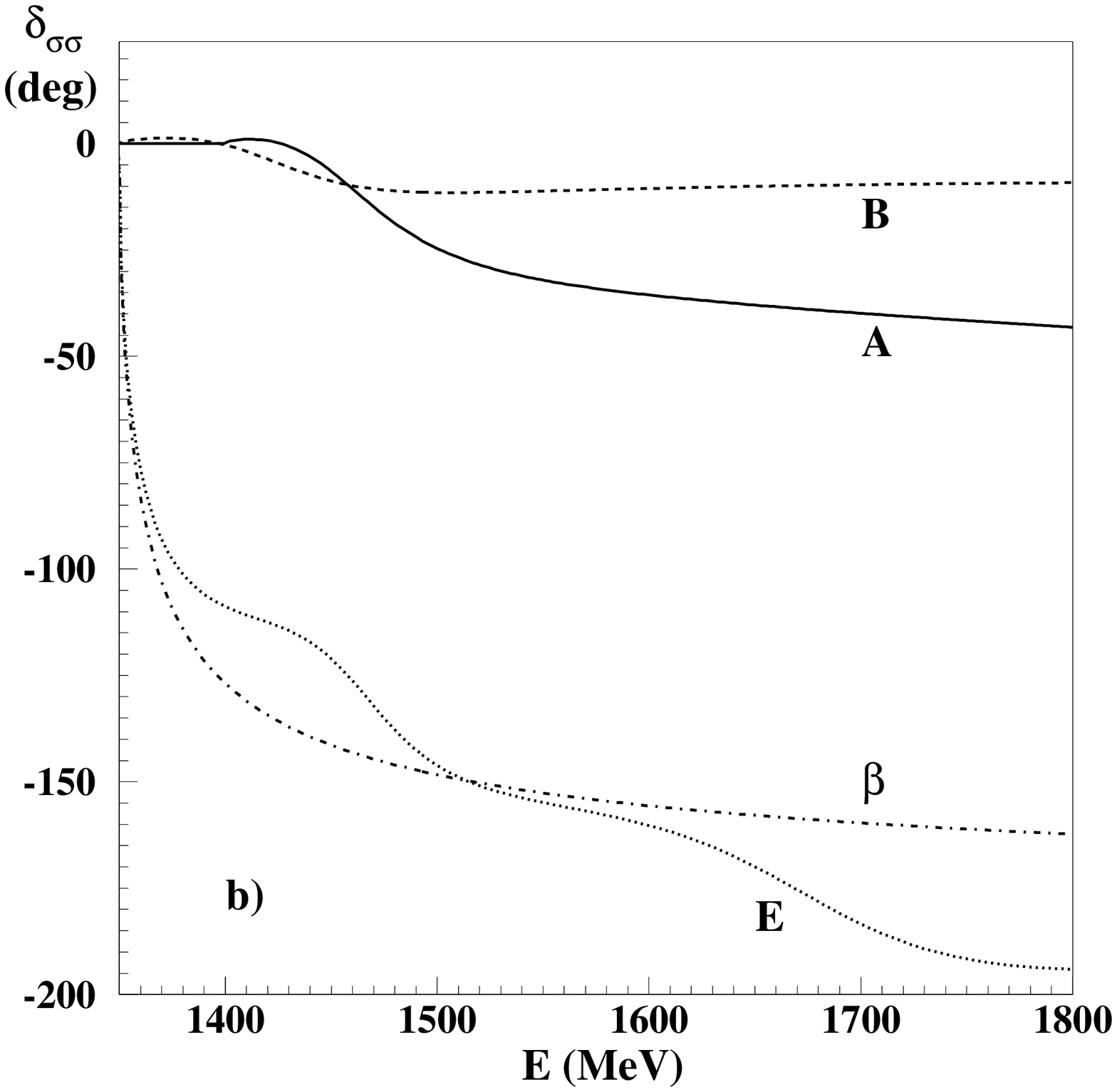}{8cm}{17}{150}{535}{650}{9cm} 
\caption{Energy dependence
for the solutions A, B and E of \roro inelasticities  {\bf a)}
and \roro phase shifts {\bf b)}. 
The curve denoted by  $\beta$ is the contribution of the double pole at 
$k_3 = -i\beta_3$ for the solution E.}
    \end{center}
                \label{abe} 
  \end{figure}
%%%%%%%%%%%%%%%%%%%%%%%%%%%%%%%%%%%%%%%%%%%%%%%%%%%%%%%%%%%%%%%%%%%%%%%

In Fig. 
%\ref{abe}a,b
3a,b 
we show inelasticities and phase shifts
in the \roro channel for 
three solutions A, B and E.
Lack of resonances above 1600 MeV for the solutions A and B is responsible for  
a smooth energy dependence of the corresponding \roro inelasticities above 1500 MeV.
A smooth decrease is also visible in the \roro phase shifts above 1450 MeV.
Influence of the \epw resonance on the \roro phase shifts can be seen 
for the solutions A and B as a small maximum around 1450 and 1400 MeV, respectively.
In the solution E a monotonous decrease is disturbed by two structures around 1425 and 1625 MeV 
related to the \epw and \fj states, respectively. The ``$\beta$'' 
curve in Fig. 
%\ref{abe}b 
3b
will be discussed in the next section. 

%%%%%%%%%%%%%%%%%%%%%%%%%%%%%%%%%%%%%%%%%%%%%%%%%%%%%%%%%%%%%%%%%%%%%%%%%%%%%%%%%%%%%%%%%%%%%%

\section{Positions of poles}

\hspace{0.7cm}
A knowledge of the positions of the $S$-matrix poles in the complex energy 
plane is important for a physical interpretation of resonances. 
Some information about the positions of poles was already given in
\cite{kll} for four solutions: A and B corresponding to the "down-flat"
data of 
\cite{klr} and C and D corresponding to the "up-flat" data.
Now we shall discuss in more detail three sets of poles, all corresponding to the 
"down-flat" data, namely the solutions A, B and E which were described in the previous section. We think that
the general structure of the solutions C and D is similar to those of sets A or B.    
In all the solutions $S$-matrix poles appear on different sheets of the complex channel 
momenta $k_1, k_2, k_3$. 
The sheets can be classified according to the signs of $Im k_1, Im k_2, Im k_3$. 
For instance, the notation $--+$ means that $Im k_1 < 0 , Im k_2 < 0$ and 
$Im k_3 > 0$. 
Positions of the most important poles for the resonances \epsig, \fo  and \epw 
were given in Table 3 of  
\cite{kll}. 
Origin of resonances can be studied by a gradual decrease of the interchannel
coupling constants. In this way, starting from the case where all interchannel
couplings are present, we arrive to the uncoupled case and obtain a trajectory
linking the  positions of a given pole from the fully coupled to the
corresponding fully uncoupled case.
The structure of poles for the various solutions is significantly different.
Positions of different $S$-matrix poles without and with couplings between 
channels are precised in Tables \ref{posa} to \ref{posf}
for solutions A, B, E and F, respectively.
In the two last columns the pole-sheet specification  and 
their labels are given. 

Let us recall that a given pole in the uncoupled channel splits into four
poles in the coupled channel case since coupling of a given channel to any other 
channel doubles the number of related poles. 
This is a consequence of the analytic structure of the Jost function $D(k_1,k_2,k_3)$
describing three coupled channels. 
Furthermore, one should remember that 
\be
D(k_1, k_2, k_3) = D^*(-k_1^*, -k_2^*, -k_3^*),
\label{symmetry}
\eb 
therefore to a given zero of $D(k_1, k_2, k_3)$ at $k_1, k_2, k_3$
there always exists a twin zero at $-k_1^*, -k_2^*, -k_3^*$. 
In the complex energy plane a pole and its twin are symmetric
with respect to the imaginary energy axis.
In an uncoupled channel case zeroes and poles of $S$-matrix lie symmetrically  
with respect to the real energy axis.
This symmetry is broken when the interchannel couplings are switched on.
Therefore asymmetry 
in the localization of the zeroes is crucial to understand the energy dependence of 
phase shifts and inelasticities in all the coupled channels.  
Now we shall discuss specific features of the different solutions.
The positions of $S$-matrix poles corresponding to the solutions A, B, E and F 
will be given in Tables 3, 4, 5 and 6, respectively. The "with couplings"
column corresponds to the fully coupled model fitted to data  while the "no
coupling" column  results  from the same solution with interchannel coupling
switched off.

\vs

{\bf\em 3a) Solution A}

\vs

For the solution A the \pp channel poles at $E = (658 - i 607)$ MeV 
and $E = (1346 - i 275)$ MeV in the uncoupled case evolve differently when the 
interchannel couplings are switched on (see Table \ref{posa}). 
%########################################################################
\begin{table}[h]
\centering
\caption{Positions of $S$-matrix poles for the solution A (in MeV)} 

\vspace{0.5cm}

\begin{tabular}{|c|c|c|c|c|c|c|}
\hline
channel & \multicolumn{2}{|c|}{no couplings} &
\multicolumn{2}{|c|}{with couplings} & Sign of & \\
\cline{2-5}
& \multicolumn{1}{|c|}{$Re E$} &
  \multicolumn{1}{|c|}{$Im E$} &
  \multicolumn{1}{|c|}{$Re E$} &
  \multicolumn{1}{|c|}{$Im E$} & $Imk_{\pi}$ $Imk_K$ $Imk_{\sigma}$ 
                               & No \\
 \hline
&&      &   564 & $-279$ & $- - -$ & I \\
&&      &   518 & $-261$ & $- + +$ & II \\
$\pi\pi$&   658 & $-607$ &   211 &    0   & $- + -$ & III \\
&&      &   532 & $-315$ & $- - +$ & IV \\
&&      &   235 &    0   & $+ + -$ & V \\
\hline
&&      &  1405 &  $-74$ & $- - -$ & VI \\
$\pi\pi$&  1346 & $-275$ &  1445 & $-116$ & $- + +$ & VII \\
&&      &  1424 &  $-94$ & $- + -$ & VIII \\
&&      &  1456 &  $-47$ & $- - +$ & IX \\
\hline
&&      &  170 &  0 & $+ - -$ & X \\
&&      &  159 &  0 & $- - -$ & XI \\
$K\overline{K}$&  881 & $-498$ &  418 &  $-10$ & $- - +$ & XII \\
&&      & 1038 &  $-204$ & $- + -$ & XIII \\
&&      &  988 &  $-31$ & $- + +$ & XIV \\
\hline
&& & 4741& $-4688$ & $- - -$ & XV \\
$\sigma\sigma$ &  118 & $-2227$ & 3687 & $-2875$& $- + -$ & XVI \\
&& & 3626 & $-3456$ & $+ - -$ & XVII \\
&& & 3533 &  $-579$ & $+ + -$ & XVIII \\
\hline
\end{tabular}
\label{posa}
\end{table}
The first pole leads to a set of poles related with 
\epsig at lower energy (poles I, II, and IV) while the second one splits into 
four states related with \epw at higher energy (poles VI to IX). 
The pole in the \kk channel at $E = (881 - i 498)$ MeV, 
lying far from the physical axis in the 
uncoupled case, moves to the \kk quasibound state at $E = (988 - i 31)$ MeV 
on sheet $-++$ (pole XIV). 
This is the \fo resonance which lies quite close to the physical axis and 
therefore it strongly influences behaviour of the \pp scattering phase 
shifts near the \kk threshold. 
The pole trajectory linking the corresponding poles in the uncoupled and coupled 
cases is drawn as a solid line in Fig. 
%\ref{trace}a.
4a.
In this figure we have also indicated at several intermediate positions
the percentage strength of the reduced interchannel couplings. 
The same broad \kk pole can also move to the pole XIII
at $E = (1038 - i 204)$ MeV on sheet $-+-$.
Its trajectory is drawn as a dotted line in Fig. 
%\ref{trace}a.
4a.
Poles III, V, X and XI, lying on the real axis of energy below the \pp threshold, appear  when
the interchannel coupling strengths are large enough. 
Their influence on the \pp phase shifts is, however, small.

The extremely wide pole
in the third channel at $E=(118 - i2227)$ MeV goes to very distant 
states at energies above 3000 MeV. 
These states, present only in solutions A and B, are quite model
dependent. They cannot be visible in the energy dependence of phase shifts or
inelasticities in the three open channels. We only give here their positions
in order to illustrate  the analytical structure of the model amplitudes. In
many phenomenological applications such distant poles are treated as background
described by additional parameters fitted separately to the data. We should
stress that in our model there is no need to introduce  any kind of artificial
background.
Large values of interchannel coupling constants (Table 1 in \cite{kll}) 
are responsible for large pole shifts as seen in Table \ref{posa} and Fig. 
%\ref{trace}a. 
4a.
This applies also to other solutions considered below, 
especially to solution B (see Tables \ref{parame}, 
\ref{posb} to \ref{posf} and Fig. 4b).
%\ref{trace}b).

%%%%%%%%%%%%%%%%%%%%%%%%%%%%%%%%%%%%%%%%%%%%%%%%%%%%%%%%%%%%%%%%%%%%%%%%%

\begin{figure}[ptb]
                \label{trace} 
    \begin{center}
\xslide{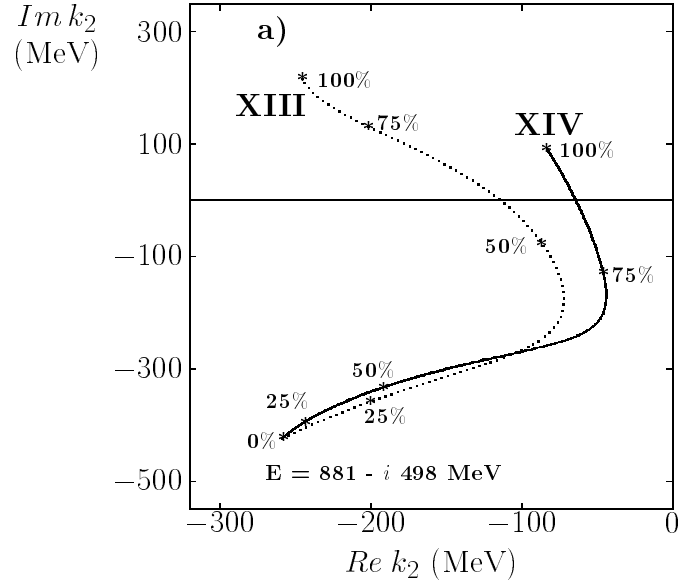}{8cm}{45}{230}{555}{705}{9cm} 

\vspace{0.5cm}

\xslide{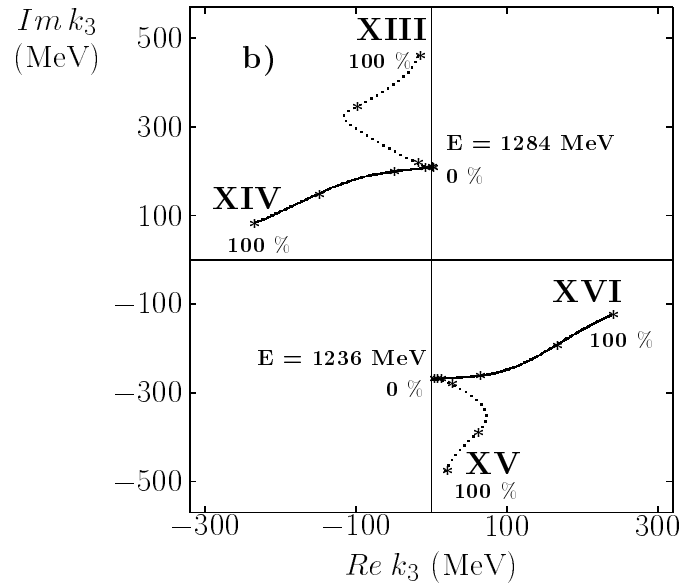}{8cm}{45}{230}{555}{705}{9cm} 
\caption{Pole trajectories in the complex momentum 
plane as a function of the percentage of the interchannel coupling strength: 
{\bf a)} for the solution A in the \kk channel, 
{\bf b)} for the solution B in the \roro channel.
Roman numbers and energies are taken from Table 3 for {\bf a)} and
from Table 4 for {\bf b)}.} 
    \end{center}
  \end{figure}

\vs

{\bf\em 3b) Solution B}

\vs
For the solution B (Table \ref{posb}) there are two wide poles 
lying far from the physical axis in the uncoupled
\pp channel. 
%###################################################################
\begin{table}[h]
\centering
\caption{Positions of $S$-matrix poles for the solution B (in MeV)} 

\vspace{0.5cm}

\begin{tabular}{|c|c|c|c|c|c|c|}
\hline
channel & \multicolumn{2}{|c|}{no couplings} &
\multicolumn{2}{|c|}{with couplings} & Sign of & \\
\cline{2-5}
& \multicolumn{1}{|c|}{$Re E$} &
  \multicolumn{1}{|c|}{$Im E$} &
  \multicolumn{1}{|c|}{$Re E$} &
  \multicolumn{1}{|c|}{$Im E$} & $Imk_{\pi}$ $Imk_K$ $Imk_{\sigma}$ 
                               & No \\
 \hline
&&       &  332 & $-114$ & $- - -$ & I\\
$\pi\pi$ &  733 & $-583$ & 511 & $-266$ & $- + +$ & II\\
&&       &  512 & $-266$ & $- + -$ & III\\
&&       &  332 & $-115$ & $- - +$ & IV\\
\hline
&&       &  900 &  $-13$ & $- - -$  & V\\
$\pi\pi$ &  999 & $-323$ & 1441 & $-125$ & $- + +$  & VI\\
&&       & 1430 & $-149$ & $- + -$  & VII\\
&&       & 942 & $-25$ & $- - +$  & VIII\\
\hline
&&       & 3670 & $-2263$ & $- - -$ & IX\\
$K\overline{K}$ &   2744  & $-1698$ & 3664    & $-2240$ & $- - +$ & X\\
&&       & 3102 &  $-940$  & $+ - -$ & XI\\
&&       & 3104 &  $-904$  & $+ - +$ & XII\\
\hline
&&  & 992  & $-34$ & $- + +$ & XIII\\
$\sigma\sigma$ &  1284 & 0 & 1421 & $-54$ & $- - +$ & XIV\\
\hline
&& & 956  &  $-36$ & $- + -$ & XV\\
$\sigma\sigma$ & 1236 & 0       & 1411 & $-85$ & $- - -$ & XVI\\
\hline
\end{tabular}
\label{posb}
\end{table}
The first pole at ($733 -i 583$) MeV splits into four states I to IV. 
Pole II on sheet $-++$ and the corresponding zero lie close to the physical axis.
Such a pole has an important influence on the \pp scattering amplitude below
1 GeV and can be related to the \epsig resonance.
The second pole evolves from $(999 - i 323)$ MeV into a pair of poles at 
$(1441 - i 125)$ and $(1430 -i 149)$ MeV both related to the \epw resonance.
The two other shifted poles V and VIII remain close to 900 MeV. 
A very broad pole at $(2744-i 1698)$ MeV in the uncoupled \kk channel leads 
to four poles IX to XII lying far away from the physical region. 
Their influence on the scattering amplitudes is negligible.
In Fig. 
%\ref{trace}b 
4b
four interesting trajectories in the $k_3$ complex plane are 
drawn. 
The \roro bound state at 1284 MeV ($Im k_3>0$) 
can either evolve to the \fo resonance (pole XIII) at $(992 - i 34)$ MeV (dotted line) 
or to the \epw (pole XIV) at $(1421 - i 54)$ MeV (solid line). 
Similarly the so-called quasibound state 
at 1236 MeV ($Im k_3 <0 $) can either go to the 
pole XV at $(956 - i 36)$ MeV closely related to the \fo
(dotted line) or to the \epw resonance (pole XVI) at 
$(1411 - i 85)$ MeV (solid line).

\vs

{\bf\em 3c) Solution E}

\vs

In Table \ref{pose} the positions of poles for the solution E are shown.
The first pole at $(542 - i307)$ MeV evolves (as for the solution B) to four wide states.
One of them (pole II) at ($533 - i 254$) MeV on sheet $-++$ lies close to the physical 
region and can be related to \epsig.
When the interchannel couplings are switched on, the second pole at $E = (1473 - i150)$ MeV,
coming from the \pp channel in the uncoupled channel case, creates four poles V to VIII.
The pole V on sheet $---$ and a zero related to the pole VI on sheet $-++$ are responsible 
for an increase of the \pp phase shifts around 1400 MeV (see Fig. 
%\ref{ephase}).
2).
Looking at inelasticities $\eta_{\pi}$ and $\eta_{K}$ in Fig. 
%\ref{etae}
1
one can see small bumps around 1400 MeV caused by the pole V.

%####################################################################################
\begin{table}[h]
\centering
\caption{Positions of $S$-matrix poles for the solution E (in MeV)} 

\vspace{0.5cm}

\begin{tabular}{|c|c|c|c|c|c|c|}
\hline
channel & \multicolumn{2}{|c|}{no couplings} &
\multicolumn{2}{|c|}{with couplings} & Sign of & \\
\cline{2-5}
& \multicolumn{1}{|c|}{$Re E$} &
  \multicolumn{1}{|c|}{$Im E$} &
  \multicolumn{1}{|c|}{$Re E$} &
  \multicolumn{1}{|c|}{$Im E$} & $Imk_{\pi}$ $Imk_K$ $Imk_{\sigma}$ 
                               & No \\
 \hline
&&       &  600 & $-355$ & $- - -$ & I\\
$\pi\pi$ &  542 & $-307$ & 533 & $-254$ & $- + +$ & II\\
&&       &  533 & $-246$ & $- + -$ & III\\
&&       &  600 & $-354$ & $- - +$ & IV\\
\hline
&&       &  1421 & $-79$ & $- - -$ & V\\
$\pi\pi$ &  1473 & $-150$ & 1441 & $-106$ & $- + +$ & VI\\
&&       &  1428 & $-104$ & $- + -$ & VII\\
&&       &  1466 & $-38$ & $- - +$ & VIII\\
\hline
&& &  328 & $-7$ & $- - -$ & IX\\
$K\overline{K}$ &  441  & 0     & 336 & $-8$ & $- - +$ & X\\
\hline
&& & 978 & $-46$ & $- + -$ & XI\\
$K\overline{K}$ & 990 & 0       & 990 & $-34$ & $- + +$ & XII\\
\hline
&&       & 1703 & $-271$ & $- - -$ & XIII\\
$\sigma\sigma$ &  1565 & $-112$ & 1648  & $-67$ & $- + -$ & XIV\\
&&       & 1673  &  $-77$ & $- - +$ & XV\\
&&       & 1624 & $-175$ & $+ - -$ & XVI\\
\hline
\end{tabular}
\label{pose}
\end{table}

The \kk antibound state at 441 MeV transforms into two states IX and X lying far away from 
the physical axes in the \kk and \roro complex momentum spaces.
The \kk bound state in the uncoupled case lies very close to the \kk threshold 
at $E=990$ MeV and can evolve, when interchannel couplings are switched on, 
into two narrow states (poles XI and XII) on sheets $- + -$ and $- + +$.
Pole XII on sheet $-++$ lies closer than pole XI to the physical region and 
therefore can be related to \fo.
The last resonance at ($1565 - i112$) MeV evolves into four states at about 1700 MeV.
As it can be seen in Fig. 
%\ref{ephase} 
2
the \pp phase shifts do not strongly 
increase around 1700 MeV.
This fact is related to the positions of two singularities of the $S$-matrix 
having the strongest influence on the \pp scattering amplitude.
One of them is the pole XIII on sheet $---$ and the second is a zero also lying on sheet $---$ 
related to the pole XVI on sheet $+--$.
The zero and the pole lie on the same sheet, so they partially cancel each other and their
influence on the energy dependence of the \pp amplitude is unusually small.
The zero lies closer to both real and imaginary energy axes  
than the pole.
Therefore in Fig. 
%\ref{ephase} 
2
a flat energy dependence of the \pp phase shifts between 
1600 and 1750 MeV is seen as well as a
smooth increase above 1750 MeV where the influence of the pole becomes larger than the action
of the zero.

Another particular case of influence of $S$-matrix singularities on scattering amplitudes 
can be seen in Fig. 
%\ref{abe}b.
3b.
A strong decrease of the \roro phase shifts for solution E, very well visible 
near threshold at 1350 MeV, can be understood if we take into account all the
singularities of the $S_{33}$ element of the $S$-matrix.
For this solution the value of parameter $\beta_3$ is very small ($\beta_3 = 92.7$ MeV)
so the double zero of $S_{33}$ at $k_3 = -i\beta_3$ is the closest singularity to the 
\roro threshold.
This zero of $S_{33}$ is a result of the Jost function double pole at
$k_3 = -i\beta_3$ as it can be seen in the analytical expression for $J_{33}$
given in Eq. (A9) of 
\cite{klm}.
One can calculate the \roro phase shifts keeping only this pole contribution 
to $J_{33}$.
The result is drawn in Fig. 
%\ref{abe}b 
3b
as the dotted line denoted by ``$\beta$``.
Comparison with the full calculation shows the dominance of this singularity over a large
energy range. 
The poles related to the \epw and \fj only slightly disturb the energy dependence of 
the \roro phase shifts at about 1425 and 1625 MeV.
This example shows that one cannot extract a complete information about
resonances from the energy dependence of phase shifts. One furthermore needs
to know the analytical structure of the $S$-matrix singularities.

Let us here note that in the Particle Data Table 
\cite{pdg98}
there is a state called $f_J(1710)$ in a mass range similar to that of our
$f_0(1710)$, with a $J=0$  possibility not excluded. 
It has been recently shown that this $J=0$ assignment is
more favorable than the $J=2$ one
\cite{kloet98}.

\vs

{\bf\em 3d) Solution F}

\vs

In the solution F there are both \kk and \roro bound states in the uncoupled
channel case (see Tables \ref{chi} and \ref{posf}). In the fully coupled
case, the \roro quasibound state with a width about 0.5 MeV generates a very
narrow $180^o$ jump of the \pp phase shifts at about 1350 MeV. 
The appearance of such a not well confirmed resonance cannot be excluded 
since the precision of the existing data is rather limited and the $\chi^2$ values for the 
solution F are still acceptable.
Apart from this narrow state, the solution F is similar to other solutions
so in the further analysis we shall not study in detail properties
of this solution.
%#####################################################################################

\begin{table}[h]
\centering
\caption{Positions of $S$-matrix poles for the solution F (in MeV)} 

\vspace{0.5cm}

\begin{tabular}{|c|c|c|c|c|c|c|}
\hline
channel & \multicolumn{2}{|c|}{no couplings} &
\multicolumn{2}{|c|}{with couplings} & Sign of & \\
\cline{2-5}
& \multicolumn{1}{|c|}{$Re E$} &
  \multicolumn{1}{|c|}{$Im E$} &
  \multicolumn{1}{|c|}{$Re E$} &
  \multicolumn{1}{|c|}{$Im E$} & $Imk_{\pi}$ $Imk_K$ $Imk_{\sigma}$
                               & No \\
 \hline
&&       &  691 & $-511$ & $- - -$ & I\\
$\pi\pi$ &  554 & $-377$ & 528 & $-255$ & $- + +$ & II\\
&&       &  673 & $-410$ & $- + -$ & III\\
&&       &  658 & $-379$ & $- - +$ & IV\\
\hline
&&       &  1387 & $-81$ & $- - -$ & V\\
$\pi\pi$ &  1407 & $-80$ & 1428 & $-93$ & $- + +$ & VI\\
&&       &  1367 & $-79$ & $- + -$ & VII\\
&&       &  1447 & $-77$ & $- - +$ & VIII\\
\hline
&& & 486 & $-5$ & $- - -$ & IX\\
$K\overline{K}$ &  497  & 0 & 396  & $-25$ & $- - +$ & X\\
\hline
&& & 967 & $-31$ & $- + -$ & XI\\
$K\overline{K}$ & 968 & 0 & 993  & $-42$   & $- + +$ & XII\\
\hline
&& & 250 & 0 & $- - -$ & XIII\\
$\sigma\sigma$  & 341 & 0  & 270 & 0 & $- + -$ & XIV\\
\hline
&&  & 1349 & $-0.2$ & $- - +$ & XV\\
$\sigma\sigma$  & 1353 &  0 & 1349 & $-0.3$ & $- + +$ & XVI\\
\hline
\end{tabular}
\label{posf}
\end{table}

\vs

Comparing different solutions we have seen that similar relatively narrow re\-so\-nan\-ces can 
emerge from very different poles in the uncoupled channel cases depending on the set 
of the interaction parameters.
Therefore one would need more data of a better precision to 
disentangle between phenomenologically good solutions A, B and E.
In the next section we shall further discuss other properties of the meson-meson 
scattering amplitudes.

%-----------------------------------------------------------------

\section{Influence of $S$-matrix poles and zeroes on phase shifts and 
inelasticities near \epw resonance} 

\hspace{0.7cm}
Properties of the scattering and production amplitudes in the energy
region near 1400 MeV can be understood provided that we know the positions of the
$S$-matrix singularities, especially the poles and zeroes close to the physical 
region.
In the previous chapter we have seen that the $S$-matrix has many poles lying 
on different sheets. 
Now we would like to choose the most important poles and zeroes which 
influence the phase shifts and inelasticities and determine the resonance 
parameters.

The $S$-matrix elements $S_{i j}\ (i, j = 1, 2, 3)$ can be 
written in terms of the Jost function of different arguments, for example

\be S_{11} = \frac{D(-k_1, k_2, k_3)}{D(k_1, k_2, k_3)}.
                                              \label{S11}\eb         
Expressions for some other matrix elements can be found in   
\cite{lles96}. 
All the $S$-matrix elements are inversely proportional
to the  Jost function $D(k_1,k_2,k_3)$ which has a zero on sheet $---$ at the
channel momenta $k_{id}$ ($i=1,2,3$):

\be    D(k_{1d},k_{2d},k_{3d}) = 0 . \label{Dzero}   \eb
The corresponding energy $E_d$ in the complex plane is given by 

\be E_d= 2 \sqrt{{k_{1d}}^2 + {m_1}^2} =2\sqrt{{k_{2d}}^2 + {m_2}^2} =
  2 \sqrt{{k_{3d}}^2 + {m_3}^2}, \label{ezero} \eb
%\be E_d= 2 \sqrt{{k_{id}}^2 + {m_i}^2}, \,\,\,\,\,\,\,\,\,i = 1,2,3 \eb 
where  $m_i$ denote meson masses. Energies $E_d$ 
on sheet $---$ corresponding to the \epw resonance are collected
in Table \ref{positions1400} for our solutions A, B, E and F. 
The knowledge of this pole position of the $S$-matrix
is, however, not sufficient to describe the pion-pion phase shifts and 
inelasticities
even at the energy closest to the pole. We should also know the {\em{zero}}   
of the numerator  $D(-k_1, k_2, k_3)$ on sheet $+++$ close to the 
physical axis. This zero is in turn  related to the zero of the denominator 
$D(k_1, k_2, k_3)$ on sheet $-++$. The corresponding energies on sheet $-++$
are also given in Table \ref{positions1400}. We can notice that both real and imaginary parts
of energy are shifted on sheet $-++$ if we compare them with the
corresponding parts of energy on sheet $---$. 
The values of phase shifts 
and inelasticities depend on the zero of the numerator 
$D(-k_1, k_2, k_3)$ and on the zero of the denominator $D(k_1, k_2, k_3)$. 
This means that 
by measurements of the \pp phase shifts and inelasticities we cannot 
uniquely determine one single value of the resonance energy and one value
of the resonance width related to the imaginary part of energy. 
Strong interchannel couplings are responsible for the energy shifts of 
zeroes found on different sheets. 
%##########################################################################
\begin{table}[h]
\centering
\caption{Energy positions of $S$-matrix poles related to $f_0(1400)$} 

\vspace{0.5cm}

\begin{tabular}{|c|c|c|c|c|}
\hline
& \multicolumn{4}{|c|}{Energy (in MeV)} \\
\cline{2-5}
Solution & sheet & sheet & sheet & sheet \\
& $- - -$ & $- + +$ & $- + -$ & $- - +$ \\
\hline
A & $1405 - i74$  & $1445 - i116$ & $1424 - i94$  & $1456 - i47$ \\  
B & $1411 - i85$  & $1441 - i125$ & $1430 - i149$ & $1421 - i54$ \\
E & $1421 - i79$  & $1441 - i106$ & $1428 - i104$ & $1466 - i38$ \\
F & $1387 - i81$  & $1428 - i93$  & $1367 - i79$  & $1447 - i77$ \\
\hline
\end{tabular}
\label{positions1400}
\end{table}

In vicinity of the \epw resonance one can, however, built up an approximation to 
the $S$-matrix elements in three channels.
This will furthermore allow us to understand the role played by the four poles 
whose energy positions in different sheets are given in Table \ref{positions1400} for
our four solutions A, B, E and F. 

We make an expansion of the Jost function $D(k_1,k_2,k_3)$ near its zero on sheet 
$---$ at $(k_{1d}, k_{2d}, k_{3d})$
\be
D(k_1,k_2,k_3) \approx (k_1- k_{1d}) d_1. \label{d}
\eb
Let us denote by $k_{in}$ ($i = 1,2,3$) the zero position of the same Jost function
on sheet $-++$. 
Then  in the first approximation 
\be
D(-k_1,k_2,k_3) \approx (k_1- k_{1n}^*) c_1. \label{c}
\eb
Here $d_1$ and $c_1$ are complex constants.
The quantity $1/d_1$ is the residuum of the $1/D(k_1,k_2,k_3)$
pole on sheet $---$ and $1/c_1$ is the residuum of the pole
on sheet $-++$.
Here we have used the property of the Jost function expressed by (\ref{symmetry}).
The twin zero can sometimes be closer than the pole on sheet $---$ to the physical 
region and therefore it can strongly influence the $S$-matrix element.
The \pp $S$-matrix element can be then approximated by
\be
S_{11} = \frac{k_1 + p_1}{k_1 - k_{1d}}f_1, \label{s11approx}
\eb
where $f_1 = c_1/d_1$ and $p_1 = -k_{1n}^*$.
Similarly the \kk and \roro $S$-matrix elements can be written as
\be
S_{22} = \frac{D(k_1, -k_2, k_3)}{D(k_1, k_2, k_3)} \approx 
\frac{k_2 + p_2}{k_2 - k_{2d}}f_2, \label{s22approx}
\eb
\be
S_{33} = \frac{D(k_1, k_2, -k_3)}{D(k_1, k_2, k_3)} \approx 
\frac{k_3 + p_3}{k_3 - k_{3d}}f_3, \label{s33approx}
\eb
where $f_2$ and $f_3$ are complex constants,
$p_2$ is the kaon complex momentum on sheet $-+-$ and $p_3$ is the $\sigma$
momentum on sheet $--+$ for which the Jost functions in numerators vanish.

We have numerically checked, by comparison with the exact
values, that the phase shifts and inelasticities for the solutions A,
B and E, in the region of the \epw resonance (effective mass range
between 1350 and 1500 MeV), are qualitatively well described by the
formulae (\ref{s11approx}) to (\ref{s33approx}). For the solution B the
agreement is even quantitative since the percentage error is
only of the order of 10$\%$ or less. 
This error is obtained if the coupling constants $f_i$ are calculated using the first order 
derivatives of the Jost function.
The errors can be reduced further if we
modify the complex factors $f_1$, $f_2$ and $f_3$ taking into account the
second derivatives of the Jost function as explained in Appendix B.
The solution E is somewhat particular since the region of
\epw is also influenced by a wide resonance \fj (see Table \ref{pose}).
The phase shifts in the \roro channel are additionally affected by the second
order zeroes at $k_3 = \pm i\beta_3$ since the parameter $\beta_3$ is small
in that case as already discussed in Section  3c.

%%%%%%%%%%%%%%%%%%%%%%%%%%%%%%%%%%%%%%%%%%%%%%%%%%%%%%%%%%%%%%%%%%%%%%%%%%%%              
\section{Limited applicability of the Breit-Wigner approach}
     
\hspace{0.7cm}              
We should note here that zeroes of the Jost function $D(k_1, k_2, k_3)$
on sheets $---$, $-++$, $-+-$ and $--+$ are in general different,
so the following inequalities between the corresponding complex momenta
hold: $p_1 \neq k_{1d} $, $p_2 \neq k_{2d} $ and  $p_3 \neq k_{3d} $.
In particular, looking at the Table \ref{positions1400} we can notice that the 
poles on sheet $--+$ are shifted towards higher energy in comparison with 
poles on sheet $---$. 
Also the corresponding width related to the imaginary part of energy 
is considerably reduced on sheet $--+$ for the solutions A, B and E.
This fact has important consequences on the energy dependence of phase
shifts and inelasticity parameters which will be different from those 
obtained using the Breit-Wigner form.

Let us now compare the approximations to the diagonal $S$-matrix elements
(\ref{s11approx}) to (\ref{s33approx}) and the Breit-Wigner multichannel
formula for the transition matrix elements

\be T_{ij}^{BW} = \frac{1}{\sqrt{k_ik_j}}\frac{M\Gamma}{M^2-s-iM\Gamma}c_ic_j.
                                              \label{TBW} \eb
In this equation $M$ is the mass and $\Gamma$  
the width of a resonance, $c_i, c_j$ are real channel branching ratios
and $c_i^2 =\Gamma_i/\Gamma$, $\Gamma_i$ being the partial
decay width. These formulae are valid if the resonance pole dominates the
transition amplitude in the physical region and if background can be
neglected. The corresponding $S$-matrix diagonal elements expressed by

\be S_{ii}^{BW}=1 + 2 i k_i T_{ii}^{BW} \label{Sdiag} \eb 
are then written as 
 
\be S_{ii}^{BW}=\frac{s-[M^2+iM(2\Gamma_i-\Gamma)]}{s-(M^2-iM\Gamma)}.
                                       \label{SdiagBW} \eb
Remembering that $s=E^2=4(k_i^2+m_i^2)$ we can approximate (\ref{SdiagBW}) by
\be S_{ii}^{BW} \approx \frac{k_{iN}}{k_{iD}}\frac{k_i-k_{iN}}{k_i-k_{iD}},
                                        \label{Sdiagk} \eb
where 
\be  {k_{iD}}^2=\frac{M^2}{4}-m_i^2 - i \frac{M\Gamma}{4},
                                        \label{kiD} \eb
and  
\be  {k_{iN}}^2=\frac{M^2}{4}-m_i^2 + i \frac{M}{4}(2\Gamma_i-\Gamma).
                                          \label{kiN} \eb
While formally (\ref{Sdiagk}) looks very similar to (\ref{s11approx}),
(\ref{s22approx})
 and (\ref{s33approx}), it is very different from them since $k_{iD}$ and $k_{iN}$ have
 to satisfy the inequalities: $Re k_{iN} \leq  Re k_{iD}$, 
 $\mid Im k_{iN}\mid \leq \mbox{$\mid Im k_{iD}\mid$}$. These inequalities follow
 from an obvious inequality that each partial width $\Gamma_i$ is smaller
 than the total width $\Gamma$. In fact each Breit-Wigner $S_{ii}$ matrix 
element depends only on three {\em real} parameters $M,\Gamma$ and $\Gamma_i$
while the matrix elements (\ref{s11approx}), (\ref{s22approx}) and (\ref{s33approx}) depend
on three {\em complex} parameters: one giving the position of the zero in the denominator,
the second giving the zero of the numerator and the third parameter being $f_i$.                               
                                                                               
The Breit-Wigner formula (\ref{TBW}) is very often used to analyze experimental data in 
order to obtain a mass and a width of different resonant states
\cite{pdg98}.
It is widely believed that it should provide the {\it same} resonance
parameters independently on the reaction channel in which the resonant
signal is detected. Our analysis, however, puts some limits on a practical
applicability of the Breit-Wigner approach. The point is that in each
reaction channel not only the $S$-matrix pole plays a significant role but
also the accompanying zero. The pole is common for all the reaction channels
but the zero is different in each channel and it cannot be
located simply by giving one number corresponding to a branching ratio or a
channel coupling constant. This point will be discussed in detail in
the two next sections. Here we can make the following remark. If one
applies the Breit-Wigner formula to the data analysis in a particular
channel then one can obtains distorted resonance parameters which are to
some extend "averaged" over the pole and accompanying zero parameters. This
effect might explain a fact that some resonant parameters "measured" in one
coupled channel can be different from those obtained in another channels if
the phenomenological model applied in the data analyses is essentially
restricted to the Breit-Wigner formula.

%%%%%%%%%%%%%%%%%%%%%%%%%%%%%%%%%%%%%%%%%%%%%%%%%%%%%%%%%%%%%%%%%%%%%

\section{Branching ratios}

\hspace{0.7cm}
Branching ratios are important parameters of hadronic resonances.
Usually they are obtained from experimental data applying the multichannel 
Breit-Wigner formula (\ref{TBW}) with some background parametrization in
a limited range of effective masses.
In our model we do not use any arbitrary background parametrization
nor the Breit-Wigner parametrization.
We fully exploit the knowledge of the analytical structure of the 
$S$-matrix. 
Some $S$-matrix poles close to the physical range can be 
related to the scalar meson  resonances as already discussed in Sections 2 and 3.
From the position of a given pole in the complex energy plane we can
deduce the mass
of the resonance and its total width.
A determination of the partial decay width in presence of two and three open
channels is, however, a more complicated issue which we are going to discuss 
in this section.

\vs

{\bf\em 6a) Definitions}

\vs

Let us recall that our model satisfies the unitarity
condition for the $S$ matrix: $S^+S = 1$ and that the
diagonal matrix elements are parametrized as
\be S_{jj} = \eta_je^{2i\delta j}, ~~j = 1,2,3, \label{sdiag} \eb
where $\eta_j$ inelasticities and $\delta_j$ the channel phase
shifts. The nondiagonal elements
are related to the non-diagonal reaction $T$-matrix elements by

\be T_{jl} = \frac{1}{2i}\frac{1}{\sqrt{k_j k_l}} S_{jl}, ~~ j, l = 1,2,3,\,\,\,\, j \neq l.
\label{tnondiag}
\eb
Expressions for $S_{jl}$ can be found in
\cite{lles96}.
The diagonal $T$-matrix elements read
\be T_{jj} = \frac{1}{2i k_j}\left(S_{jj} - 1\right), \,\,\,\,\, j = 1,2,3. \label{tdiag} \eb
$T$-matrix elements satisfy the unitarity equations.
For example, in the first channel
above the \roro threshold
\be Im T_{11} = k_1|T_{11}|^2 + k_2|T_{12}|^2 + k_3|T_{13}|^2.\eb
The total cross section in the first channels reads:
\be \sigma^{tot}_{11} = {\frac{8\pi}{k_1}} Im T_{11} .   \label{sigtot}      \eb
The elastic cross section is given by
\be \sigma^{el}_{11}= 8\pi|T_{11}|^2                   \label{sigel}       \eb
and the transition cross sections from the channel 1 to the channels $j=2$ or 3
are expressed by 
\be \sigma_{1j}= \frac{8 \pi}{k_1} k_j |T_{1j}|^2 .       \label{sig1j}    \eb
These cross sections satisfy the equation: 
\be \sigma^{tot}_{11} =  \sigma^{el}_{11}+\sigma_{12}+\sigma_{13}.  \label{sum} \eb

In Fig. 
%\ref{sigma} 
5
all the \pp cross sections corresponding to the solution B are shown.
At low energies we see a huge peak with a maximum near 600 MeV. It can be
attributed to a very wide scalar $\sigma$ meson. Near 1 GeV one notices
a very deep and narrow minimum which is related to the \fo meson. The fact
that this resonance is seen as a dip and not, as in the most cases as, a maximum
of the total cross section, is due to the special value of the elastic \pp phase
shift which goes through $180^o$ slightly below 1 GeV. 
If it happens below the \kk threshold the total cross section is equal to 0. 
The next minimum
of the total cross section is near 1500 MeV where another resonance \fgg 
appears. Up to the \kk threshold the \pp scattering is elastic but above
the \kk threshold the \pp to \kk transition cross section becomes visible.
The transition to the $\sigma\sigma$ channel is clearly visible above
the third threshold at about 1350 MeV.
%%%%%%%%%%%%%%%%%%%%%%%%%%%%%%%%%%%%%%%%%%%%%%%%%%%%%%%%%%%%%%%%%%%%
\begin{figure}[ptb]
                \label{sigma} 
    \begin{center}
\xslide{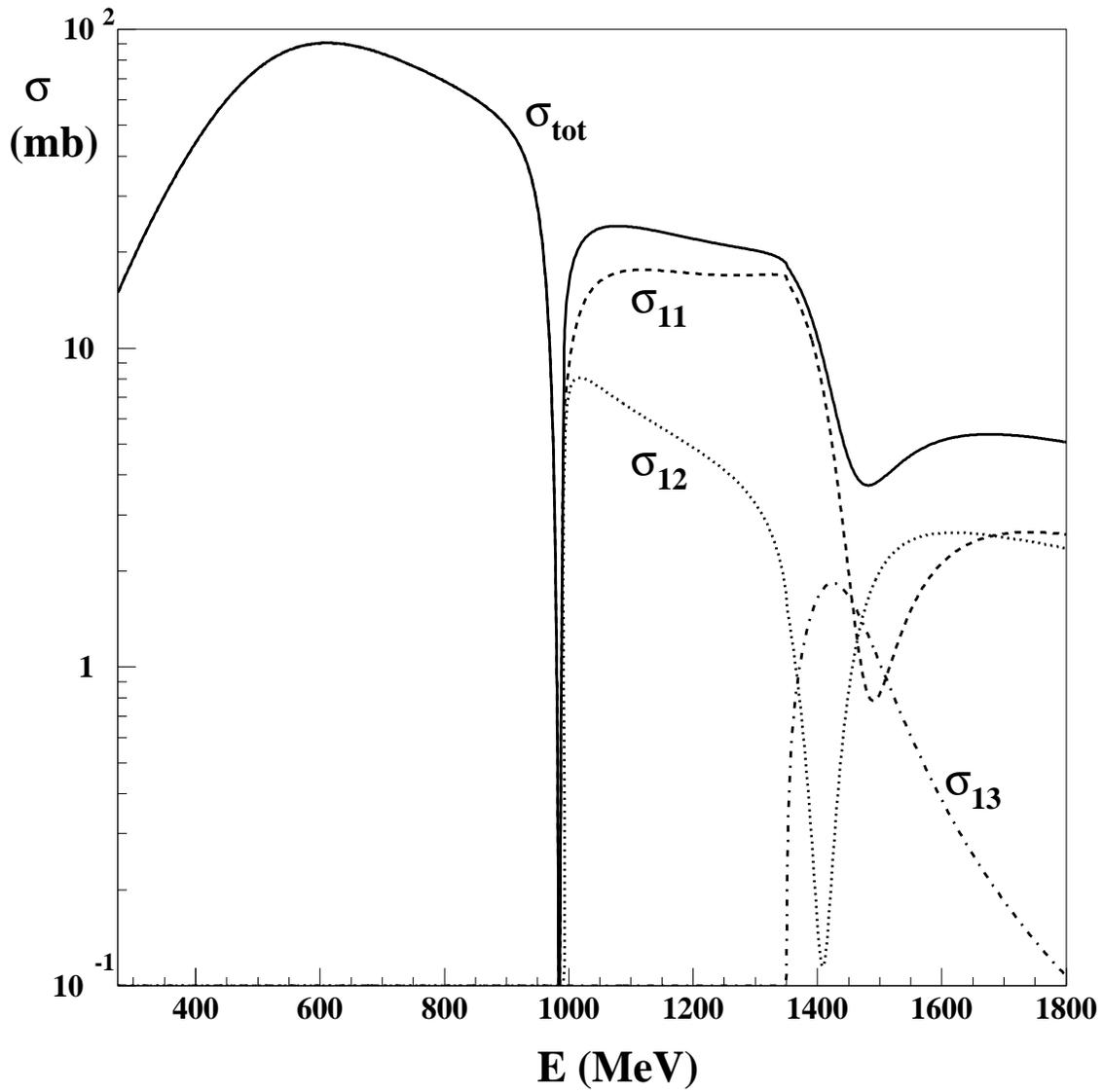}{15cm}{20}{145}{535}{655}{15cm} 
\caption{Energy dependence of cross sections $\sigma_{ij}$ for the solution B} 
    \end{center}
  \end{figure}

Numbers of meson pairs produced in each channel are proportional
to the cross sections defined above so for example in the \pp channel
we can define three branching ratios
\be b_{1j}=\frac{\sigma_{1j}}{\sigma^{tot}_{11}}~, ~~j = 1,2,3.\label{branch} \eb
These branching ratios obviously satisfy
\be b_{11}+b_{12}+b_{13}=1.                                \label{bsum}   \eb
Cross sections similar to $\sigma_{1j}$ and the corresponding branching ratios
can be defined in channels 2 and 3 by changing index 1 into 2 and 3,
respectively. In this way
one can obtain nine branching ratios $b_{ij}$ ($i,j=1,2,3$). All these 
quantities as well as the cross sections are functions of energy or the
effective mass. Below the third threshold where energy is smaller than $2m_3$
but greater than $2m_2$
the branching ratio matrix $b_{ij}$ reduces to 2x2 matrix containing only
four nonzero matrix elements $b_{11},b_{12},b_{21}$ and $b_{22}$.
If the $T$-matrix is approximated by the Breit-Wigner formula (10) then
the branching ratios are given simply by 

\begin{equation}
b_{ij}^{BW} = c_j^2 .
\label{bijcj}
\end{equation}

From (\ref{bijcj}) we infer that the corresponding branching ratio matrix contains
identical rows.
\vs

{\bf\em 6b) Discussion}

\vs

We shall discuss in detail the behaviour of branching ratios 
for the solution B, plotted in Fig. 
%\ref{br}.
6.
Energy dependence of branching ratios for solutions A and E
is similar to that for solution B.   
We do not expect that the energy behaviour of the branching ratios 
for the "up-flat" solutions (solutions C and D in \cite{kll})
will be qualitatively different from that of the "down-flat" solutions
A and B.

At first let us discuss the behaviour of the "elastic" branching ratio  $b_{11}$
as a function of energy in the pion-pion channel (see Fig. 
%\ref{br}a). 
6a).
At the \kk threshold $b_{11}=1$ but it decreases 
enormously steeply with increasing energy to about 0.6 (solution A and E),
0.55 (solution B) or 0.5 (2-channel "down-flat" solution of 
\cite{kll}) 
already at the energy
close to 1 GeV. This steep decrease is related to the opening of the \kk channel.
Next $b_{11}$ steadily increases to a maximum close to 0.9 for the above
solutions at the energy reaching about 1.4 GeV. At this energy the third
threshold is open for the solutions A, B and E. For $E>$1.4 GeV $b_{11}$ decreases
very fast to a minimum close to 0.2 for three cases A, B and E and to 0 for
2-channel "down-flat" solution. Then it rises again especially steeply for
the latter solution and slower for the solutions A and B. This minimum is
closely related to the presence of $f_0(1400)$. 

%%%%%%%%%%%%%%%%%%%%%%%%%%%%%%%%%%%%%%%%%%%%%%%%%%%%%%%%%%%%%%%%%%%%%%%%%%%

\begin{figure}[ptb]
                \label{br} 
    \begin{center}
\xslide{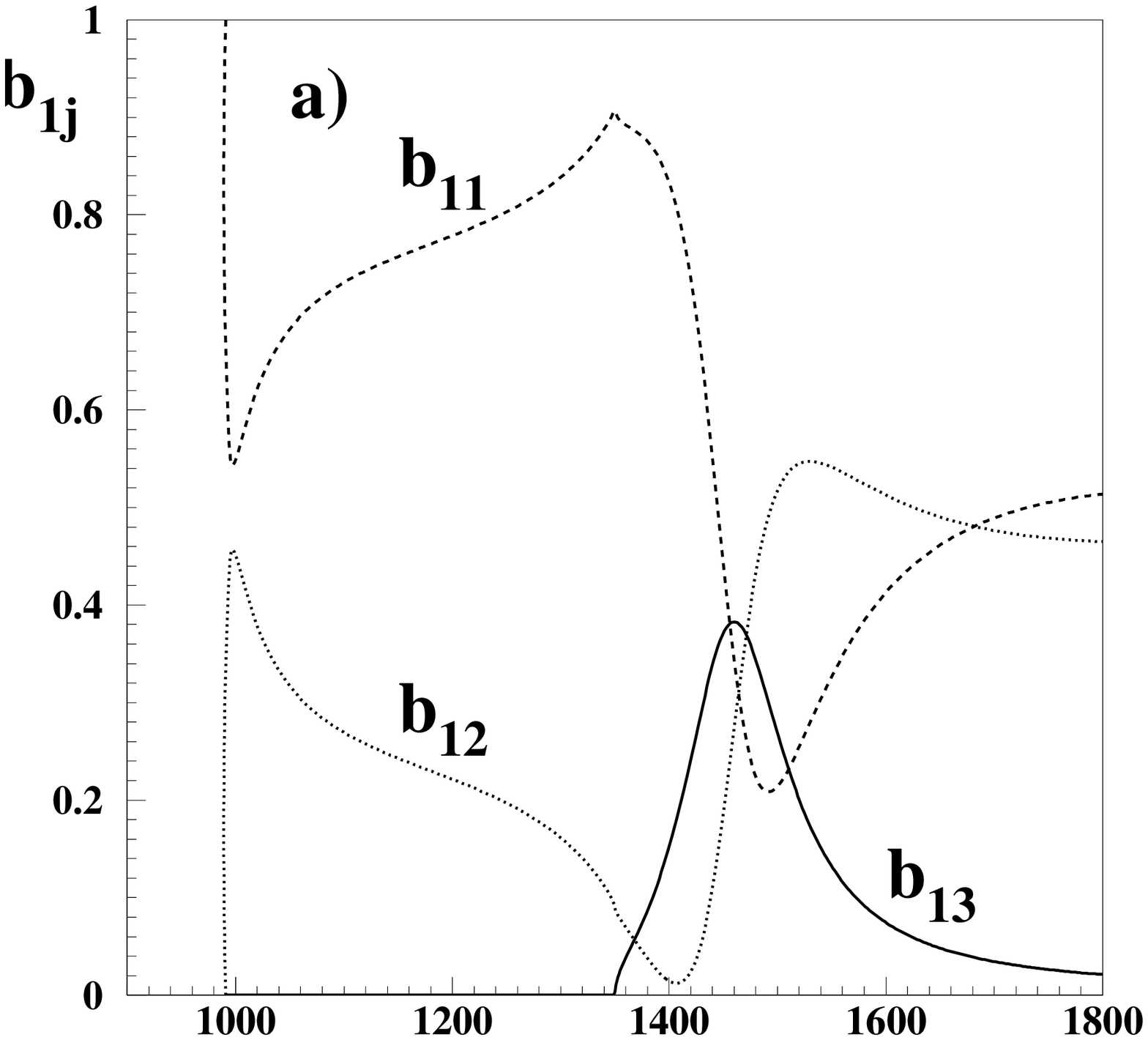}{6cm}{16}{170}{540}{650}{9cm} 
\xslide{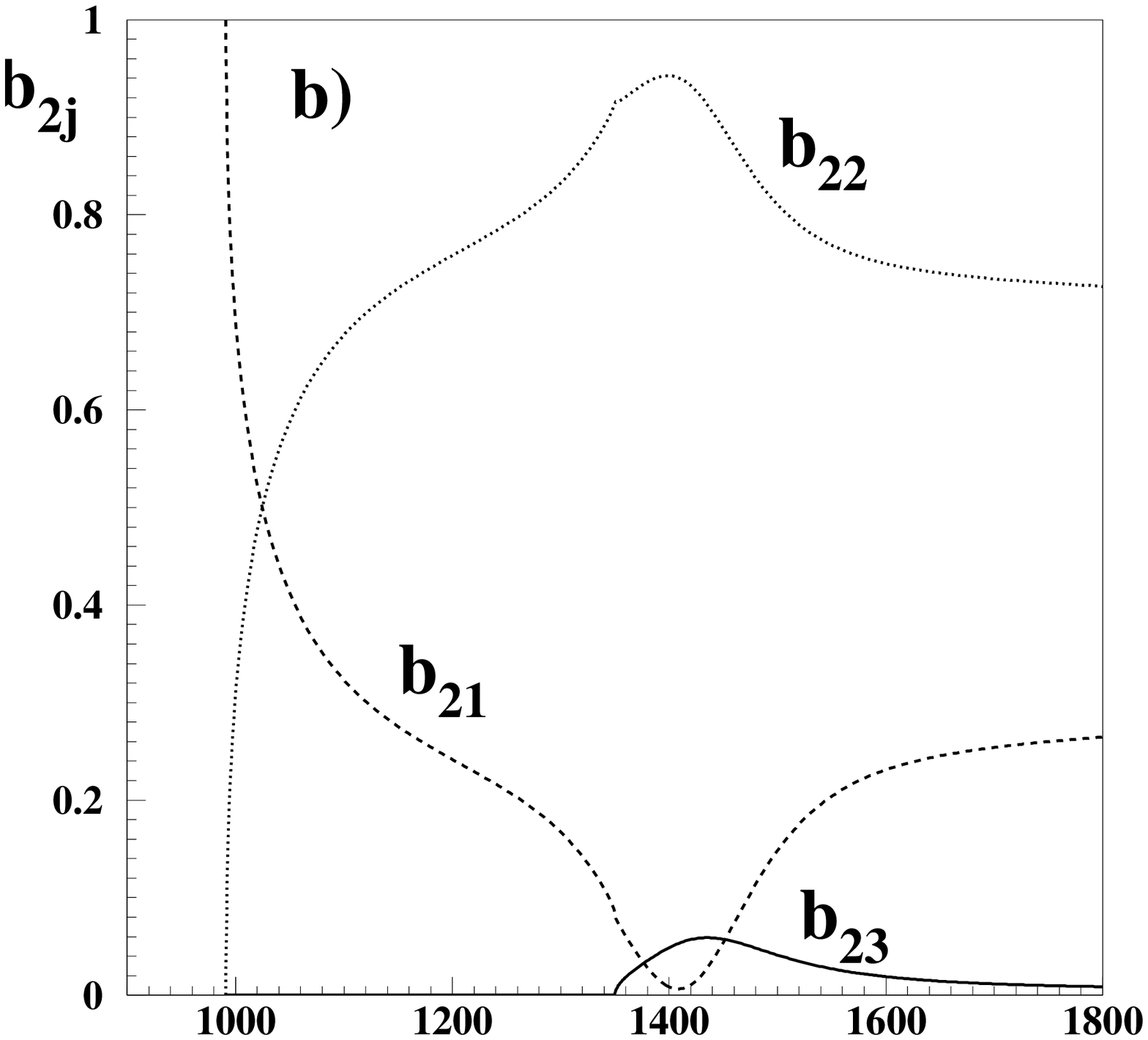}{6cm}{16}{170}{540}{650}{9cm} 
\xslide{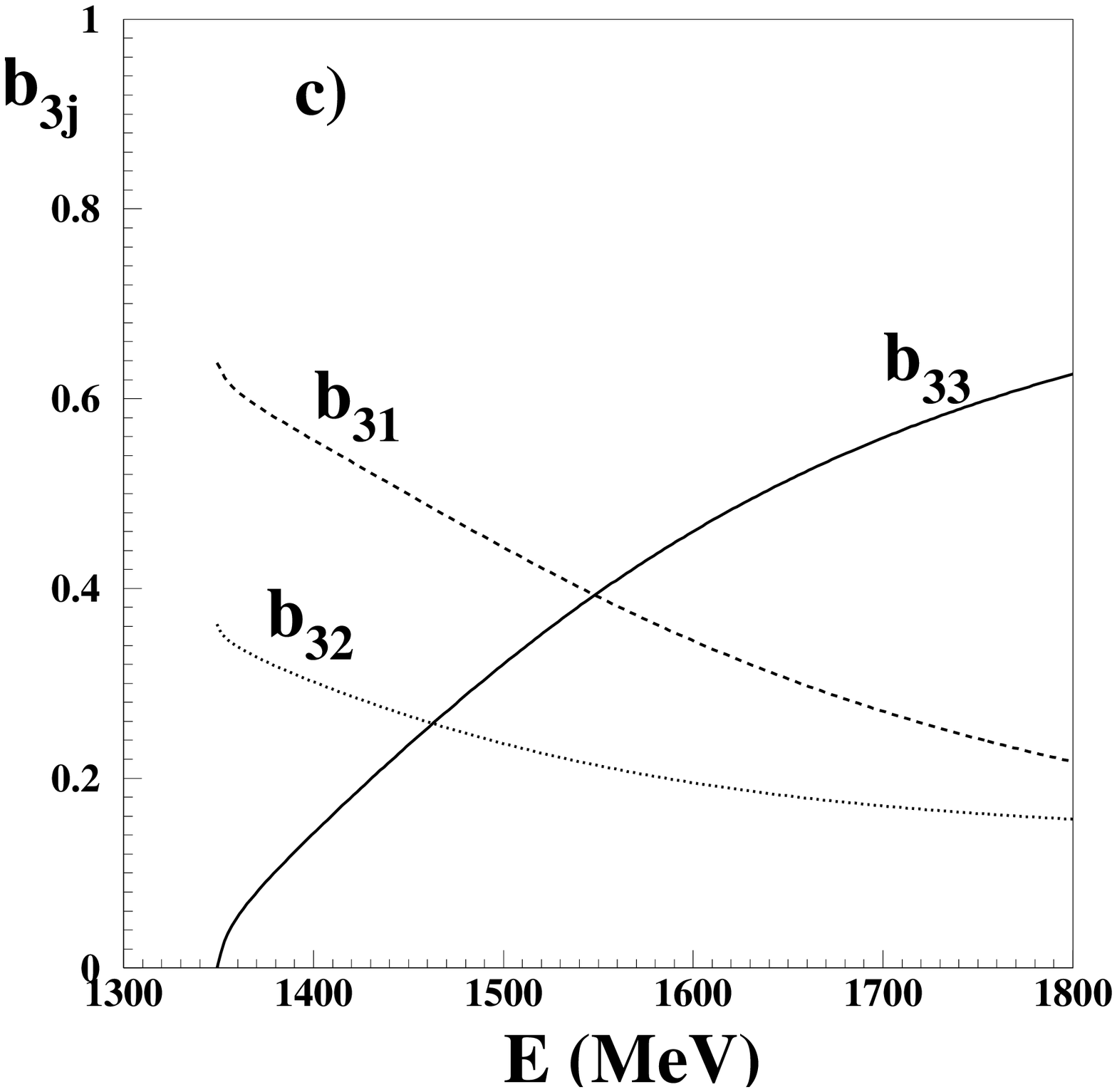}{6.2cm}{16}{140}{540}{650}{9cm} 
\caption{Energy dependence of branching ratios $b_{ij}$ for the
solution B.
Dashed line corresponds to $j$ = 1, dotted line to $j$ = 2 and 
solid line to \mbox{$j$ = 3.} 
In {\bf a)} $i$ = 1 (\pp channel), in {\bf b)} $i$ = 2 (\kk channel) and in fig. 
{\bf c)} \mbox{$i$ = 3} (\roro channel).} 
    \end{center}
  \end{figure}

The $b_{12}$ branching ratio is equivalent to the transition probability
of two pions into a pair of two kaons. Below the third threshold ($E<2 m_3$)
$b_{12}=1-b_{11}$, so its behaviour is completely determined by $b_{11}$.
The $b_{12}$ rises very steeply from the \kk threshold attaining a maximum
near 1 GeV, then decreases to a minimum value at about 1.4 GeV (see Fig. 
%\ref{br}a). 
6a).
At this
energy, however, the third threshold opens and the transition to the $\sigma
\sigma$ channel becomes possible as stated by (\ref{bsum}). An effect
of the interplay between three channels is that  $b_{12}$ rises above 1.4 GeV 
attaining a maximum and then decreases rather slowly with energy for $E>$1.5 GeV.

Above the third threshold in the \pp and \kk channels a nonzero fraction
of the channel total cross section originates from the transition to the
$\sigma\sigma$ state. In the \pp channel the branching ratio  $b_{13}$
forms a regular maximum at $E=1.48$ GeV, 1.46 GeV and 1.48 GeV for the solutions  
A, B and E, respectively.
The width of this peak is about 110 to 120 MeV (see
Fig. 
%\ref{br}a). 
6a).
The position and the width of this peak corresponds very closely
to the parameters determined for the total width \fgg by the Crystal Barrel 
Collaboration in different channels 
\cite{pdg98}.

In the second channel the $b_{21}$ element describes the ratio of the 
\kk to \pp transition cross section to the total \kk cross section. 
At the \kk threshold the annihilation cross section $\sigma_{21}$ tends to 
infinity and the elastic cross section is finite, therefore the $b_{21}$ 
coefficient is equal to 1. 
With increasing energy it decreases to a quite deep minimum 
close to 0 at about 1.4 GeV. At higher energy it rises slowly again for all
three solutions. The transition element $b_{23}$ starts from 0
at the $\sigma\sigma$ threshold, reaches rather quickly a maximum and then
drops smoothly (see Fig. 
%\ref{br}b).
6b).

Two branching ratios in the third channel, namely $b_{31}$ and $b_{32}$,
decrease from the large values at the $\sigma\sigma$ threshold. Their
behaviour is characteristic for the annihilation reactions. On contrary,
the elastic branching ratio $b_{33}$ rises monotonically with energy as shown
in Fig. 
%\ref{br}c.
6c.
Comparing Figs. 6a, 6b and 6c one can see that the Breit-Wigner formulae (\ref{bijcj})
are never satisfied above 1 GeV even in vicinity of the well defined resonances
like $f_0(980)$ and $f_0(1400)$.

\vs

{\bf\em 6c) Average branching ratios near the \fo resonance}

\vs

The \fo resonance lies very close to the \kk threshold and this fact has
a very important consequence on the experimental determination of the partial
widths of this resonance for the decay into \pp and \kk channels. The \pp
channel is open below the \kk threshold so for this channel we can determine
the averaged \pp branching ratio $b_{11}$ choosing a range of the \pp 
effective mass ranges centered at the resonance mass $M_s$ with a
maximum energy value equal to $M_{max}$. This value should be larger
than $M_s$ by more than the resonance width. This width is 
still not well defined experimentally. 
In   \cite{pdg98} the full width is 40 to 100 MeV.
In our three channel analyses \cite{kll} it was between 60 and 70 MeV
and larger than these numbers by 10 to 30 MeV for two channel fits. 
Therefore we can choose the value $M_{max}=1100$ MeV as the upper limit in the
integration of the \pp or \kk effective mass distributions. The average \pp
branching ratio is thus defined as follows:
\be <b_{11}> = \frac{1}{2(M_{max}-M_s)} \int_{2M_s-M_{max}}^{M_{max}} b_{11}(E)
 dE.
                                                           \label{bav11}    \eb 
The average \kk branching ratio over the same energy interval is 
\be  <b_{12}> = 1 - <b_{11}> .                                \label{bav12}     \eb 
Since $b_{12} =0$ below the \kk threshold 
\be <b_{12}> = \frac{1}{2(M_{max}-M_s)} \int_{2 m_2}^{M_{max}}
b_{12}(E) dE.
                                                        \label{bav12int}  \eb                                                         
The mass values $M_s$ corresponding to the \fo resonance position on sheet
$-++$ are 993, 989, 992 and 990 MeV for two channel down-flat model and 
three channel models A, B and E, respectively \cite{kll}. After integration 
over energy we have obtained averaged $<b_{12}>$ values of 0.191, 0.156, 0.170 
and 0.158 for 
the above four solutions corresponding to different sets of 
interaction parameters. 
Above numbers should not be, however, compared with the experimental
value $21.9\pm 2.4\%$ quoted as $\Gamma(K\overline{K})/\left[ 
\Gamma(\pi\pi) + \Gamma(K\overline{K})\right]$ in the previous editions of
the Reviev of Particle Pro\-per\-ties
(see for example 
\cite{pdg9294}).
The reason is that the numbers written under the title 
$"\Gamma_{\pi\pi}/(\Gamma_{\pi\pi} + \Gamma_{KK})"$ in 
\cite{pdg9294,pdg98} 
are the values of inelasticity coefficient $\eta_{av}$, defined below,  and therefore
they are {\em not} equal to the partial branching ratios of \fo.
This difference is important since the branching ratios defined by
(\ref{branch}) depend not only on the inelasticity coefficient $\eta$ but also on the 
\pp and \kk phase shifts.

Experimentally 
the $\eta_{av}$ value was obtained from the data on the 
$\pi^- p \longrightarrow K_S^0 K_S^0 n$ reaction using the relation:
\be
\eta^2(E) = 1- \frac{\sigma(\pi^+\pi^- \longrightarrow f_0(980) 
\longrightarrow K_S^0 K_S^0)}{\sigma_u},
\label{etasquared}
\eb
where $\sigma_u = \pi/(6k^2)$ is the unitary limit of the 
$\pi^+\pi^- \rightarrow K_S^0 K_S^0$
scalar isoscalar cross section 
with $k$ being the $K_S^0$ momentum in the $K_S^0 K_S^0$ c.m. system.
Let us remark that from the unitarity condition, below opening of the third channel, one has 
$\eta \equiv \eta_1 \equiv \eta_2$.  
Thus one can define
\be
\eta_{av} = \left( \frac{1}{E_{max}-2m_K}\int_{2m_K}^{E_{max}} dE\; \eta^2(E) \right)^{1/2} 
\label{etaav}
\eb
and use $E_{max} \approx 1.1$ GeV as a good representation of the upper 
experimental limit of the $K_S^0 K_S^0$ effective energy
(see  
\cite{loverre}
for a discussion of experimental uncertainties).
Experimental values of $\eta_{av}$ obtained in
\cite{loverre, wetzel, cason}
are equal to $0.67 \pm 0.09$, $0.78\pm 0.03$ and $0.81^{+0.09}_{-0.04}$
respectively.     
Our calculations give {\em lower} values: 
0.487, 0.496 and
0.494 for the solutions A, B and E, respectively.
We have also calculated $\eta_{av}$ values 0.494 and 0.761 
corresponding to the two-channel "down-flat" solution of   
\cite{kll}
and to the two-channel set 1 result of   
\cite{klm}, respectively.
Here we should stress that the last number is a result of the fit to the data 
for the reactions $\pi^- p \longrightarrow K^- K^+ n$ and 
$\pi^+ n \longrightarrow K^- K^+ p$
of  
\cite{cohen}
while the four previous numbers around 0.49 are based on inelasticities measured in the 
\reactpol 
\cite{klr}.
The difference between above two-channel fits is therefore due to use of 
different experimental data sets.
Problems with normalization of the $\pi\pi \rightarrow K\overline{K}$
cross sections have already been discussed by Morgan and Pennington in 
\cite{morgan}
(see in particular Fig. 4 therein) and
Bugg, Sarantsev and Zou in
\cite{bugg96}.
The values of inelasticity $\eta$ from 
\cite{klr},
which were later used in
\cite{kll},
correspond to the elastic $\pi\pi \rightarrow \pi\pi$ reaction.
Although the errors are quite large the $\eta$ values 
found in
\cite{klr}
near the
\fo resonance are visibly lower then those of  
\cite{cohen}.  
This difference, however, could be partially explained by a possible contribution of channels
other than \pp and \kk. 

This difference between the  data can also be seen in the calculation of
$<b_{12}>$.
If we use the parameters corresponding to the set 1 of   \cite{klm}
then $<b_{12}>$=0.055.  This low value is related to rather low mass of \fo 
equal to 973 MeV and to its width 29 MeV obtained in 
\cite{klm}.
In the fits described in 
\cite{kll}
we have obtained the \fo masses close to 990 MeV and substantially larger values of 
the \fo width.
Authors of 
\cite{klm} and \cite{cdl} have also improperly used other definition of "branching ratio" in
their Eq. (52) and in Eqs. (33) to (35), respectively.
Those equations gave essentially the averaged values of $1-\eta^2$ which should not be 
directly compared to $\eta_{av}$.

One remark is in order here. If all the transition amplitudes are dominated
by a single Breit-Wigner resonance as in (\ref{TBW}) then the branching
ratio 2x2 matrix has to satisfy the following relations: $b_{21}=b_{11}$
and $b_{22}=b_{12}$. 
These relations are not satisfied above the \fo resonance
energy (see Fig. 
%\ref{br}).
6).
 It means among others that the transition amplitudes cannot be
simply described in terms of a single Breit-Wigner formula.

In Fig. 
%\ref{br}a,b 
6a,b
we see that the element $b_{21}$ is in general larger than 
$b_{12}$
especially near the \kk threshold where $b_{21}$ tends to 1 while
$b_{12}$ goes to zero. Nevertheless there is rather wide range 
of energies between 1.15 and 1.4 GeV where  both ratios are quite
close each other.       

\vs

{\bf\em 6d) Average branching ratios in the range 1100 - 1420 MeV}

\vs

In \cite{gorlich} the mass range 1100 to 1420 MeV has been chosen to compare 
the ratio of \pp to \kk pairs produced by incoming pions on a polarized target
at about 18 GeV/c. The \kk branching ratio was $6.4^{+1.6}_{-2.0}$ \%. 

We have
calculated the average values of $b_{12}$ in this energy region obtaining
0.189, 0.175, 0.166 and 0.167 for the two-channel "down-flat" and 
three-channel solutions A, B and E, respectively. These values are higher than
the experimental result of   \cite{gorlich}.
One should, however,  remember that the experimental errors of phase shifts and inelasticities
are large (see
\cite{kll}) 
so the theoretical errors of $b_{12}$ are also large.
For completeness we give the value
0.099 corresponding to the data set 1 of   \cite{klm}. In order to get 
this value and the other four values written above we used the ordinary
averaging procedure 
 
\be \overline b_{12}= \frac{1}{M_{max}-M_{min}} \int_{M_{min}}^{M_{max}}
 b_{12}(E) dE                                                                                             
                                              \label{nbav12} \eb 
and not the definition (\ref{bav12int}) which is applicable only near the
\kk threshold. 
Here one should once again remind the confusion concerning the definitions
of the branching ratios as discussed in subsection 6c, namely the value
$B_{av} = 16 \pm 1\%$ calculated in
\cite{klm}
should not have been compared with the experimental branching ratio
$6.4^{+1.6}_{-2.0}\%$.  
                                             
\vs

{\bf\em 6e) Average branching ratios near the \epw resonance}

\vs

In presence of three open channels the branching ratio matrix {\boldmath
$b$} has 9 elements. As already discussed in 6b, 
behaviour of these elements in three different channels \pp, \kk and \roro
is shown in Fig. 
%\ref{br} 
6
for the solution B. For the solutions A and E the general
shape of curves is quite similar although numerical values of various
branching ratios differ. In \cite{kll} we have found that the width of the
scalar resonance called here \fgg varies between 90 and 180 MeV depending on
the solution. Now we can choose the energy interval 1350 MeV to 1500 MeV to
present the averaged values of the branching ratios in the form of 3x3
matrix. The elements of the last line of this matrix are averaged over the
energies larger than the third threshold energy equal to $2 m_3$. For the
solutions A, B and E we obtain

\vspace{0.9cm} 
 
\hspace{3cm}{\boldmath $\bar b$}$_A$ =

\vspace{-1.4cm}
                                                 
 \[ \left(  \begin{array}{ccc}
                0.636 & 0.127 & 0.237    \\
                0.061 & 0.844 & 0.095    \\
                0.300 & 0.407 & 0.293     
          \end{array}           \right) \] 
          
%For the solution B the corresponding matrix reads:

\vspace{0.9cm}

\hspace{3cm}{\boldmath $\bar b$}$_B$ = 

\vspace{-1.4cm}

\[ \left( \begin{array}{ccc}
                0.608 & 0.163 & 0.229    \\
                0.054 & 0.900 & 0.045    \\ 
                0.529 & 0.286 & 0.185     
          \end{array}             \right) \] 
          
\vspace{0.9cm} 

and \hspace{2.2cm}{\boldmath $\bar b$}$_E$ =

\vspace{-1.4cm}

\[ \left( \begin{array}{ccc}
                0.604 & 0.133 & 0.263    \\
                0.063 & 0.813 & 0.124    \\
                0.120 & 0.156 & 0.724     
          \end{array}             \right) \]
In the \kk channel
$\overline{b}_{22}$ dominates over $\overline{b}_{21}$ and $\overline{b}_{23}$. 
Here the probabilities of the 
\kk to \pp or \roro transitions are quite small. In the \roro channel, however,
for the solutions A and B there are strong transitions from \roro to \pp or \kk 
transitions which actually are comparable to the transitions from the \pp channel to 
 \roro channels.

In   \cite{amsler97} the branching ratios for the $f_0(1500)$ decay into
five channels, \pp, $\eta\eta$, $\eta\eta^{,}$, \kk and $4\pi$, are given as
29, 5, 1, 3 and 62 \%, respectively. The two main
disintegration channels are \pp and $4\pi$. In this model the $4\pi$ channel
is represented by the effective
\roro channel and we also obtain large fractions for the  averaged branching
 ratios $\overline{b}_{11}$ and $\overline{b}_{13}$. If we calculate the ratios
 $b_{13}/b_{11}$ exactly at 1500 MeV then we obtain numbers 
 2.4, 1.2 and 2.3 for the solutions A, B and E, respectively. These numbers illustrate the
 importance of the $4\pi$ channel in agreement with experimental result of
 \cite{amsler97}. In \cite{abele98} the ratio $r = (B[f_0 \rightarrow
 K\overline{K}]/B[f_0 \rightarrow \pi\pi])\,k_1/k_2 = 0.24 \pm 0.09$
 is calculated (here by $f_0$ we mean $f_0(1500)$). If we define the ratio
 $\overline{b}_{12}/\overline{b}_{11}$ then we obtain values 0.20, 0.27 and 0.22 for the solutions
 A, B and E, respectively. These values are close to $r$. From the partial
 decay widths of the $f_0(1500)$ given in \cite{bugg96} one can calculate
 $\Gamma_{K\overline{K}}/\Gamma_{\pi\pi} \approx 0.10 \pm 0.05$ which is
 smaller than $r$ but still consistent within the experimental errors. We
 know, however, that extraction of the branching ratios from experiment is a
 difficult task as it is, for example discussed in \cite{abele96bis}. We
 should mention, however, that the average branching ratios depend quite
 sensitively on the energy bin chosen in the actual calculation as it can be
 seen in Fig. 
 %\ref{br}a. 
 6a.
 Furthermore we see that the branching ratio
 $b_{12}$, corresponding to the \pp$\rightarrow$\kk transition, is very
 small around 1420 MeV, close to the position of our \epw resonance poles.
 This is in qualitative agreement with the small number for the \kk
 branching ratio (3 \%) given in \cite{amsler97}.

%%%%%%%%%%%%%%%%%%%%%%%%%%%%%%%%%%%%%%%%%%%%%%%%%%%%%%%%%%%%%%%%%%%%%%%%%%%%%%%%
\section{Coupling Constants}

\hspace{0.7cm}
Coupling constants are useful physical quantities related to the 
residuum of the $S$-matrix at its complex pole value
\be
s_R = M^2 - iM\Gamma, 
\label{residuum}
\eb
where  $M$ is a resonance mass and $\Gamma$  its total width.
We define the coupling constants $g_i$ by the formula 
\be \frac{g_ig_j}{4\pi} = i\sqrt{s_R} \lim_{s\to s_R} \left[\left(s -
s_R\right)\frac{S_{ij}(s)}{\sqrt{k_ik_j}}\right]. \label{gigj}
\eb
The product of the constants $c_ic_j$ appearing in the Breit-Wigner formula
(\ref{TBW}) is proportional to $g_ig_j$:
\be c_i c_j = \frac{[(s_R/4 - m_i^2)(s_R/4
- m_j^2)]^{1/4}}{2M\Gamma \sqrt{s_R}} \frac{g_ig_j}{4\pi}. \label{cicj}
\eb
In this way the diagonal Breit-Wigner coupling constant corresponding to the 
matrix element $S_{ii}$ given by (\ref{SdiagBW}) reads
\be
\frac{g_{iBW}^2}{4\pi} = 2M\Gamma_i\frac{\sqrt{s_R}}{k_{iD}}, \label{giBW}
\eb
where $k_{iD}$ is defined by (\ref{kiD}).
One can notice that the knowledge of $g_i^2/4\pi$ in the Breit-Wigner approach is 
equivalent to a determination of the partial width $\Gamma_i$.
The two other independent quantities are $M$ and $\Gamma$.

The coupling constants corresponding to the approximations
(\ref{s11approx} to \ref{s33approx}) are:
\be
\frac{g_i^2}{4\pi} = 8i\sqrt{s_R}f_i(k_{id}+p_i).
\label{gi4pi}
\eb
They are complex and depend on six parameters in contrast to (\ref{giBW}) 
where only three parameters appear.
 
Using (\ref{gigj}) we have calculated the values of the coupling constants  
for three solutions A, B and E at two resonances \fo and \epw. 
They are shown in Tables \ref{gifo} and \ref{giepw}, respectively.
We can notice that the dispersion of the values 
$\mid g_1\mid^2/4\pi$ corresponding to the \pp channel is the smallest one since 
we have  mostly fitted the \pp data.
The worst situation is in the \roro channel where there are no data available.
In Table \ref{giepw} we can notice that the \kk coupling constants at the 
\epw resonance are much smaller than the \pp coupling constants
contrary to the \fo resonance as seen in Table \ref{gifo}.
The knowledge of the coupling constants is not sufficient to describe fully 
the $S$-matrix elements since the coupling constants are only related
to the poles. 
We need also positions of the zeroes as discussed in Sect. 4.
%###########################################################################
\begin{table}[h]
\centering
\caption{Coupling constants of \fo at the pole $-++$ (in GeV$^2$)}

\vspace{0.5cm}

\begin{tabular}{|c|c|c|}
\hline
& & \\
Solution & $\frac{\mid g_1 \mid^2}{4\pi}$  & $\frac{\mid g_2 \mid^2}{4\pi}$  \\
& & \\
\hline
A & 0.37 & 1.84 \\
B & 0.41 & 1.33 \\
E & 0.40 & 1.94 \\
\hline
\end{tabular}
\label{gifo}
\end{table}
\begin{table}[h]
\centering
\caption{Coupling constants of \epw at the pole $---$ (in GeV$^2$)}

\vspace{0.5cm}

\begin{tabular}{|c|c|c|c|}
\hline
& & & \\
Solution & $\frac{\mid g_1 \mid^2}{4\pi}$  & $\frac{\mid g_2 \mid^2}{4\pi}$ & 
$\frac{\mid g_3 \mid^2}{4\pi}$ \\
& & & \\
\hline
A & 0.580 & 0.106 & 0.091 \\
B & 0.607 & 0.233 & 0.433 \\
E & 0.579 & 0.107 & 0.217 \\
\hline
\end{tabular}
\label{giepw}
\end{table}

%%%%%%%%%%%%%%%%%%%%%%%%%%%%%%%%%%%%%%%%%%%%%%%%%%%%%%%%%%%%%%%%%

\section{Phenomenological parametrizations \\ of multichannel amplitudes}

\hspace{0.7cm}
Following our studies described in previous chapters we shall here make some
proposal for a simple parametrization of multichannel amplitudes. This
parametrization, based on existence of poles and zeroes of the $S$-matrix,
has {\em limited} energy range of applicability. If there is some evidence
of the presence of a well separated resonance
in some energy region one can use expressions (\ref{s11approx}) to
(\ref{s33approx}) to describe the diagonal $S$-matrix
elements. The corresponding diagonal $T$-matrix elements (\ref{tdiag}) read:
\be
T_{ii} = \frac{1}{k_i}(f_i - 1 + f_i\frac{k_{id} + p_i}{k_i-k_{id}}).
\label{Tii}
\eb

From (\ref{sdiag}) one can obtain the phase shifts and inelasticities
in each channel provided that the three complex parameters ($k_{id}, p_i$
and $f_i$), determined in a fit to some data, satisfy the unitarity
conditions $\mid S_i \mid^2 \leq 1$ in a given energy range. The
off-diagonal $S$-matrix elements (\ref{tnondiag}) can then be directly
calculated as explained in
\cite{lles96}
for the three channel case.

The fitted  parameters $k_{id}$, related in all channels by the energy conservation
(\ref{ezero}), provide us with the resonance energy position and its width.
Other complex parameters $p_i$ are, however, not related by (\ref{ezero}) since 
they correspond to zeroes on different sheets.
If we consider a three-channel fit and if the resonance appears well above the third threshold
then the pole lies on sheet $---$ which means that $Im k_{id} < 0$.
The effective parameters $f_i$ will take into account the second order derivative
corrections to the approximated Jost function as explained in Appendix B
(see in particular equations (\ref{sapp}) to (\ref{fk1a})).

In Sect. 5 we have discussed the Breit-Wigner approximation.
Our formula (\ref{Tii}) can be reduced to the Breit-Wigner one if $f_i=1$
and $Re k_{id} \approx - Re p_i$.
Dy\-na\-mi\-ca\-lly $f_i=1$ means that the derivatives of the Jost functions in numerators and
denominators of the $S_{ii}$ matrix elements (like in (\ref{S11})) are identically the same 
(see for example (\ref{f1})).
The phenomenology related to the formula (\ref{Tii}) is richer than 
that corresponding to the Breit-Wigner multichannel one (\ref{TBW})
since it depends on six real parameters while the latter one depends only on three
real parameters in each channel.
If one adds the so-called elastic background to the Breit-Wigner amplitude
$T_{ii}^{BW}$ then one gets the four-parameter formula
\be
T_{ii} = \frac{e^{2i\delta_B}-1}{2ik_i} + e^{2i\delta_B}T_{ii}^{BW},
\label{TiiBW}
\eb
where $\delta_B$ is a background phase.
Even introducing an inelasticity $\eta_B$ to the background amplitude like
\be
T_{ii} = \frac{\eta_Be^{2i\delta_B}-1}{2ik_i} + \eta_Be^{2i\delta_B}T_{ii}^{BW}
\label{TiiBW2}
\eb
leads to a five-parameter formula. It is only when one enlarges the number
of independent parameters in the $T^{BW}_{ij}$ (\ref{TBW}), by allowing
$c_i$ to be complex, that (\ref{TiiBW2}) is essentially
equivalent to our approximation (\ref{Tii}). 
The complex residues for broad overlapping states
have been already introduced by several authors, in particular in 
\cite{au}.
Limitations of the
Breit-Wigner formula have been already pointed out in previous sections.
Here we see that a knowledge of the $S$-matrix poles is not sufficient to
construct the scattering amplitudes. We must also know positions of the
nearby zeroes $S$-matrix in order to describe the data with sufficient
accuracy.

A case in which one encounters more than a single resonance lying near the
threshold should be treated differently especially if one wishes to study
the energy range containing that threshold. Then in general more than one
$S$-matrix pole should be taken into account. In our analysis near 1400 MeV
such a secondary pole appears on sheet $--+$ close to the physical region.
Therefore this pole is included in Table \ref{resonances} showing average
values of masses and widths of scalar resonances \mbox{\epsig,} \fo and \epw
obtained by us. Its mass is higher and its width is smaller than those of the
$---$ pole. 
The three states $f_0(500)$, $f_0(980)$ and $f_0(1400)$ can be regarded as the 
model-independent resonances since they appear in all our solutions.  Positions
of other far lying poles above 3 GeV for the solutions A or B, of the pole 
around 1700 MeV in solution the E and of the pole at 1349 MeV in solution F  
are strongly model dependent.

%####################################################################################

\begin{table}[h]
\centering
\caption{Average masses and widths of resonances \epsig, \fo and 
\epw found in our solutions A, B, E and F. Here errors represent 
the maximum departure from the average.} 

\vspace{0.5cm}

\begin{tabular}{|c|c|c|c|}
\hline
resonance & mass (MeV) & width (MeV) & sheet \\
\hline 
\epsig or $\sigma$ & $523 \pm 12$ & $518 \pm 14$ & $-++$ \\
\hline
\fo & $991 \pm 3$ & $71 \pm 14$ &  $-++$ \\
\hline
& $1406 \pm 19$ & $160 \pm 12$ & $---$ \\
\epw & $1447  \pm 27$ & $108 \pm 46$ & $--+$ \\
\hline 
%\fj & 1703 & 542 & $---$ \\
% & 1624 & 350 & $+--$ \\
%\hline
\end{tabular}
\label{resonances}
\end{table}

Ending this chapter let us remark that the formula (\ref{Tii}) can be used to approximate
the meson-meson transition amplitudes in different reactions where meson pairs appear.
This would be the case in the meson production processes by hadrons, leptons and photons, in $\gamma\gamma$ 
reactions or in the central $pp$ production, in heavy particle decays, in baryon-antibaryon 
annihilation and so on.  

%%%%%%%%%%%%%%%%%%%%%%%%%%%%%%%%%%%%%%%%%%%%%%%%%%%%%%%%%%%%%%%%%

\section{Summary}

\hspace{0.7cm}
We have presented two new solutions E and F resulting from fits to
experimental data on \pp and \kk phase shifts and inelasticities. 
Both new
solutions are characterized by the existence of a \kk bound state when the
interchannel couplings are switched off. 
In the previously found solutions
A and B the \kk pair remained unbound in the uncoupled case (see Table \ref{chi}).
In Tables \ref{posa} to \ref{posf} we have given for each solution the set of 
S-matrix poles lying on different sheets.
Knowledge of their positions is necessary in
order to evaluate masses, widths, branching ratios and coupling constants
of scalar resonances studied here. 
We furthermore stress that not only
poles of the S-matrix but also its {\it zeroes} play a very important role
in evaluation of the resonance parameters. 
All zeroes are dynamically
related. Also the knowledge of the pole trajectories as function of the
interchannel coupling strengths is important to find origin of resonances.

The fact that we have obtained different solutions while fitting the 
experimental data is mainly due to the limited number of data and due to
their limited precision. We hope, however, that this situation can be improved
in future when new data of better precision appear and the same model is
applied to  a wider class of reactions. Then one should be able to get nearly
model  independent information on the resonances lying sufficiently close to the
physical energy axis. However, the poles lying far away from this axis will
obviously be a subject of some model dependence.

We have paid a special attention to the phenomenological analysis of the
energy region near 1400 MeV where new scalar resonances appear 
\cite{pdg98}.
A simple model of the S-matrix in three channels: \pp, \kk and 4$\pi$
represented by an effective \roro has been constructed ((\ref{s11approx}) to
(\ref{s33approx}) and (\ref{Tii})). This parametrization could be used to
represent phenomenologically the meson-meson amplitudes in different
reactions. The model is based on the knowledge of the pole on sheet $---$
common to all three channels and the zero which is specific for each
channel. The diagonal $S$-matrix elements depend on three complex parameters
(positions of the pole and of the zero, and the strength). This is in
contrast to the Breit-Wigner formalism where only three real parameters
describe resonance amplitudes. The Breit-Wigner approach has limited
applicability since it restricts too much the positions of zeroes and thus
it does not take fully into account the interchannel dynamics.

Branching ratios for different channels were defined and calculated for two-
and three- channel models. We have studied the energy regions of the \fo and
\epw resonances and the region between them. We have noticed that different
definitions of branching ratios were used in literature 
which lead to some confusion in past. The result is that there are no
data on the \pp and \kk branching ratios of \fo and the incorrect
determinations of them were no more given in the new edition of the Particle
Data Tables
\cite{pdg98}.

Above the third threshold the branching ratios form a
$3\times3$ matrix. Each row of this matrix describes three branching ratios
in the particular channel, but only two of them are independent quantities
(\ref{bsum}). Phenomenological information contained in such matrix is much
richer than in the Breit-Wigner approach where only three branching ratios
can be used. We have studied energy interval 1350 MeV to 1500 MeV and found
small branching ratios corresponding to the transitions from the \pp to \kk
channel and vice versa. This is in qualitative agreement with findings of
the Crystal Barrel Group
\cite{amsler97, abele98}.

We have given formulae (\ref{gigj}), (\ref{giBW}) and (\ref{gi4pi}) for the
coupling constants in the three-channel model and its different
approximations. Calculated values of the coupling constants in the full
model are reported in Tables
\ref{gifo} and \ref{giepw}.
The \fo couples strongly to \kk and the coupling of \epw is much stronger
for the \pp channel than for the \kk channel. One should point out that the
knowledge of the resonance mass and width and the coupling constants is not
sufficient to give a good phenomenological description of the meson-meson
dynamics.

We have found that a fit of the \pp data of 
\cite{klr} together with \kk phase shifts of 
\cite{cohen} does require a $f_0(1406 \pm 19)$ on sheet $---$ and a
$f_0(1447 \pm 27)$ on sheet $--+$ (see Table 10). This seems to indicate that
these data are quite compatible with Crystal-Barrel and other LEAR
experiments which needs a broad $f_0(1370)$ and a narrower $f_0(1500)$ 
\cite{pdg98}.
To make a more closer relation to the results obtained by these experiments
one could perform a simultaneous analysis of the $\pi^-p_{\uparrow}
\rightarrow \pi^+ \pi^- n$ data and the data on
$p\overline{p} \rightarrow 3\pi^0, 5\pi^0$ and
other reactions. The model described by us can provide the production and
scattering amplitudes of 9 reactions. These amplitudes can be applied in  the
description of the final state interactions between particles produced in
different reactions. If needed our model of meson-meson interactions can be
extended to describe more than three coupled channels (like $\eta\eta$,
$\eta\eta'$ and so on).
Our model is also sufficiently flexible to accommodate scalar mesons at
energies higher than 1500 MeV, like the $f_J(1710)$ (see Table \ref{pose}).

In the present version of our model the mass of the third channel is
introduced as a free parameter. It is possible to construct a more involved
model where one smears out this mass over a rather broad range according to
the observed invariant \pp mass distribution. Crossing
symmetry constraints, if implemented, might also lead to reduce the number
of parameters. Evidently new measurements should be performed
near different meson-meson thresholds since even at the \kk threshold the
present data do not allow to obtain the \fo branching ratios. New data would
help us to test different models of inter-meson interactions. It will
lead us to a more profound knowledge of the structure of scalar mesons.

\vspace{1cm}

{\em Acknowledgments}

We would like to thank Krzysztof Rybicki for very useful discussions.
This work has been performed in the framework of the IN2P3--Polish laboratories Convention
(project No 93-71).   

%%%%%%%%%%%%%%%%%%%%%%%%%%%%%%%%%%%%%%%%%%%%%%%%%%%%%%%%%%%%%%%%%%%%%%%%%%%%%%%%%%%%
\renewcommand{\thesection}{}
\setcounter{equation}{0}
\renewcommand{\theequation}{A\arabic{equation}}
\section{Appendix A}

\hspace{0.7cm}
Here we give the full formula for the Jost function $D \equiv D(k_1,k_2,k_3)$
of our unitary model built to describe the three channels:
\pp, \kk and the effective 2$\pi$2$\pi$ (\roro). Using the same  notation 
as in (\cite{kll}) we have

\be 
D = d_0D_0 + \lambda_{00}I_{01}D_1 -
\lambda_{02}I_{22}D_2 -\lambda_{03}I_{33}D_3,
\label{ddef}
\eb  
where 
\begin{eqnarray}
d_i & = & 1-\lambda_{ii}I_{ii},\,\,\,\,\,\,\,\,\,\,\,\,\,\,\,\,\,\,\,\,\,\,i = 0,1,2,3, \label{di}\\
D_0 & = & \widetilde{D_0} - \lambda_{02}I_{01}I_{22}C_2 - \lambda_{03}I_{01}I_{33}C_3,
\label{d0} \\
C_2 & = & \lambda_{12} + I_{33}(\lambda_{13}\lambda_{23}-\lambda_{12}\lambda_{33}),
\label{c2}\\
C_3 & = & \lambda_{13} + I_{22}(\lambda_{12}\lambda_{23}-\lambda_{13}\lambda_{22}),
\label{c3}\\
\widetilde{D_0} & = & d_1d_2d_3 - 2\lambda_{12}\lambda_{13}\lambda_{23}I_{11}I_{22}I_{33} \nonumber\\
&& -\lambda_{13}^2I_{11}I_{33}d_2 -\lambda_{12}^2I_{11}I_{22}d_3 -
  \lambda_{23}^2I_{22}I_{33}d_1 \label{d0bar}, \\
D_1 & = & - I_{01}d_2C_1 + \lambda_{23}(\lambda_{11}\lambda_{23}-
 \lambda_{12}\lambda_{13})I_{01}I_{22}I_{33} \nonumber \\
&& - (\lambda_{02}I_{00} + \lambda_{12}I_{01})I_{22}C_2 - \lambda_{03}I_{00}I_{33}C_3,
 \label{D1} \\
C_{1} & = & \lambda_{11} + I_{33}(\lambda_{13}^2 - \lambda_{11}\lambda_{33}), \\
D_i & = & \lambda_{0i}\{I_{01}^2[\lambda_{11} + I_{jj}(\lambda_{1j}^2 - 
\lambda_{11}\lambda_{jj})] + I_{00}[d_1d_j - \lambda_{1j}^2I_{11}I_{jj}]\} \nonumber \\
&& + \lambda_{0j}I_{jj}[\lambda_{23}I_{00}d_1 + \lambda_{11}\lambda_{23}I_{01}^2
+ \lambda_{12}\lambda_{13}(I_{00}I_{11} - I_{01}^2)] \nonumber \\
&& + \lambda_{1i}I_{01}d_j + \lambda_{1j}\lambda_{23}I_{01}I_{jj}, \label{Di}
\end{eqnarray}
where $i=2,3$, $j=3$ if $i=2$ and $j=2$ if $i=3$.

%%%%%%%%%%%%%%%%%%%%%%%%%%%%%%%%%%%%%%%%%%%%%%%%%%%%%%%%%%%%%%%%%%%%%%%%%%%%%%

\setcounter{equation}{0}
\renewcommand{\theequation}{B\arabic{equation}}

\section{Appendix B}
%Approximation to $S_{11}$ element}

\hspace{0.7cm}
We derive an approximated formula for the pion-pion $S$-matrix element given by 
(\ref{S11}). 
At first let us make an expansion
of the Jost function in the denominator of $S_{11}$ near its zero on sheet $---$
at $\bfk_d=(k_{1d}, k_{2d},k_{3d})$:

\be D(\bfk) \equiv D(k_1, k_2, k_3) \approx (k_1-k_{1d}) d_1 + \frac{1}{2}(k_1-k_{1d})^2 s_1.
                              \label{Dapp} \eb

Here

\be d_1=\left[\frac{\partial D(k_1, k_2, k_3)}{\partial k_1}  
+\frac{k_{1d}}{k_{2d}}\frac{\partial D(k_1, k_2, k_3)}{\partial k_2}
+\frac{k_{1d}}{k_{3d}}\frac{\partial D(k_1, k_2, k_3)}{\partial k_3}\right]_{\bfk = \bfk_d}
                                     \label{cde} \eb
and
                                     
\ba  s_1 & = & \left[\frac{\partial^2 D(k_1, k_2, k_3)}{\partial k_1^2}                                   
+\frac{{k_{1d}}^2}{{k_{2d}}^2} \frac{\partial^2 D(k_1, k_2, k_3)}
                              {\partial k_2^2}
+\frac{{k_{1d}}^2}{{k_{3d}}^2} \frac{\partial^2 D(k_1, k_2, k_3)}
                              {\partial k_3^2}\right.
                                   \nonumber \\
& + & \left.2 \frac{k_{1d}}{k_{2d}} \frac{\partial^2 D}{\partial k_1 k_2} 
+ 2 \frac{k_{1d}}{k_{3d}} \frac{\partial^2 D}{\partial k_1 k_3} 
+ 2 \frac{{k_{1d}}^2}{k_{2d} k_{3d}} 
\frac{\partial^2 D}{\partial k_2 k_3}\right]_{\bfk = \bfk_d}. \label{de} \ea

In (\ref{cde}) and (\ref{de}) we have used relations

\be k_2 - k_{2d} \approx (k_1-k_{1d}) \frac{k_{1d}}{k_{2d}},  \eb
 
\be k_3 - k_{3d} \approx (k_1-k_{1d}) \frac{k_{1d}}{k_{3d}},  \eb                                        
which follow from the energy conservation (\ref{ezero}).
Similarly we make a power expansion of the Jost function in the numerator 
of $S_{11}$. 

Let us denote by 
$\bfk_n=(k_{1n}, k_{2n}, k_{3n})$ a zero of $D(k_1, k_2, k_3)$ on sheet $-++$:

\be    D(k_{1n},k_{2n},k_{3n}) = 0 . \label{seczero}   \eb

Then using (\ref{symmetry}) we expand

\be D(-k_1, k_2, k_3) \approx (k_1-{k_{1n}}^ {*}) c_1 
+ \frac{1}{2}(k_1-{k_{1n}}^{*})^2 t_1,
                                     \label{Dnumapp} \eb
where 
                                    
\ba c_1 & = & \left[-\frac{\partial D(-k_1, k_2, k_3)}{\partial k_1} 
-\frac{k_{1n}}{k_{2n}}\frac{\partial D(-k_1, k_2, k_3)}{\partial k_2}\right.
                         \nonumber \\
& - & \left.\frac{k_{1n}}{k_{3n}}\frac{\partial D(-k_1, k_2, k_3)}
                         {\partial k_3}\right]_{\bfk = \bf-k^*_n} 
                                     \label{cn} \ea
and $t_1$ is given by an equation similar to that for $s_1$ in (\ref{de}).

With the help of (\ref{Dapp}) and (\ref{Dnumapp}) the
pion-pion $S$-matrix element reads

\be S_{11}\approx\frac{k_1-k_{1n}^*}{k_1-k_{1d}} f_1 ,   \label{sapp} \eb
where 

\be f_1=\frac{c_1}{d_1} F(k_1), \label{f1} \eb

\be F(k_1)=\frac{1+\frac{1}{2}\frac{t_1}{c_1}(k_1-{k_{1n}}^{*})} 
                            {1+\frac{1}{2}\frac{s_1}{d_1}(k_1-k_{1d})} .
                            \label{fk1} \eb
The function  $F(k_1)$ can further be approximated by a constant if we limit
ourselves to values $k_1$ close to $Rek_{1d}$ in the denominator of $F(k_1)$ and 
close to $Rek_{1n}$ in its numerator:

\be F(k_1)\approx \frac{1+\frac{1}{2}\frac{t_1}{c_1} i Im k_{1n}}
                            {1-\frac{1}{2}\frac{s_1}{d_1}i Im k_{1d}} .
                            \label{fk1a} \eb  
                            
Similar expressions can be derived for the diagonal \kk and \roro $S$-matrix elements.                                           
                            
%%%%%%%%%%%%%%%%%%%%%%%%%%%%%%%%%%%%%%%%%%%%%%%%%%%%%%%%%%%%%%%%%%%%%%%%%%%%%%%%%%%%%%%%%%%%%%%%%
%-----------------------------------------------------------------

%\include{bibliography}

\end{document}